\documentclass[a4paper,12pt]{article}
\usepackage{amsmath}
\usepackage{amsfonts}
\usepackage{amssymb}
\usepackage{latexsym}
\usepackage{epsfig}
\usepackage{graphicx}
\usepackage{oldgerm}
\usepackage{theorem}
\usepackage{bm}
\usepackage{enumitem}

\usepackage[hang,small,bf]{caption}
\usepackage[subrefformat=parens]{subcaption}
\def\N{\mathbb{N}}
\def\C{\mathbb{C}}
\def\R{\mathbb{R}}

\def\X{\bm{X}}
\def\Y{\bm{Y}}
\def\Z{\bm{Z}}
\def\x{\bm{x}}

\def\cP{\mathcal{P}}
\def\cT{\mathcal{T}}

\def\sm{\mathsf{m}}

\def\Var{\mathrm{Var}}

\def\bra{\langle}
\def\ket{\rangle}

\theorembodyfont{\itshape}

\begin{document}

\title{\bf 
Weighted Point Configurations \\ with Hyperuniformity: \\ An Ecological Example and Models
}
\author{
Ayana Ezoe,
\footnote{
Department of Physics,
Faculty of Science and Engineering,
Chuo University,
Kasuga, Bunkyo-ku, Tokyo 112-8551, Japan
} \,
Makoto Katori,
\footnote{
Department of Physics,
Faculty of Science and Engineering,
Chuo University,
Kasuga, Bunkyo-ku, Tokyo 112-8551, Japan;
e-mail:
makoto.katori.mathphys@gmail.com
} \,
and
Tomoyuki Shirai,
\footnote{
Institute of Mathematics for Industry,
Kyushu University,
744 Motooka, Nishi-ku,
Fukuoka 819-0395, Japan;
e-mail: shirai@imi.kyushu-u.ac.jp
}
}
\date{24 May 2025}
\pagestyle{plain}
\maketitle

\begin{abstract}
Random point configurations are said to be 
in hyperuniform states, if density fluctuations are
anomalously suppressed in large-scale.
Typical examples are found in Coulomb gas systems
in two dimensions especially called log-gases in
random matrix theory, in which
points are repulsively correlated by long-range potentials.
In infertile lands like deserts continuous survival 
competitions for water and nutrition will cause long-ranged
repulsive interactions among plants.
We have prepared digital data of spatial configurations
of center-of-masses for bushes weighted by 
bush sizes which we call masses.
Data analysis shows that such ecological point configurations
do not show hyperuniformity as unmarked point processes,
but are in hyperuniform states as marked point processes
in which mass distributions are taken into account.
We propose the non-equilibrium statistical-mechanics models 
to generate marked point 
processes having hyperuniformity, in which 
iterations of random thinning of points and coalescing
of masses transform initial uncorrelated point processes
into non-trivial point processes with hyperuniformity.
Combination of data analysis 
and computer simulations
shows the importance of strong correlations 
in probability law
between spatial point configurations 
and mass distributions of individual points 
to realize hyperuniform marked point processes.
\end{abstract}

\section{Introduction}
\label{sec:introduction}
Structures and distribution of ordered and disordered 
configurations of points in a space have been important research subjects in physics, 
where the points represent the locations of
atoms or molecules in gases, liquids, glasses, quasicrystals,  
and crystals. 
If we consider a classical ideal-gas, the spatial configurations
of molecules are completely random and 
the statistical ensembles of such uncorrelated
point configurations is identified with
the \textit{Poisson point process} (PPP)
studied in mathematics \cite{DVJ03,BB10}.
In mathematical physics and probability theory 
spatial statistics of (random) configurations of points are
simply called a (random) \textit{point process} 
in general \cite{DVJ03,BB10}. 
Correlated point processes have been also
extensively studied in statistical mechanics
for Coulomb gas systems, which are especially called
the \textit{log-gases} in one- and two-dimensional
spaces \cite{For10}. 
It is known that they are realized as
the eigenvalue distributions of 
Hermitian and non-Hermitian random matrices \cite{For10,Gin65,CL95,Meh04,Kat16}
and as the zeros of random analytic 
functions \cite{HKPV09,KS22}.

Recently in condensed matter physics and related material sciences,
correlated particle systems are said to be in a
\textit{hyperuniform state}, 
when density fluctuations are anomalously suppressed in
large-scale \cite{Tor18}.
Consider a bounded domain with linear size $\ell$ in the $d$-dimensional space,
which we call an observation \textit{window}.
In this window we measure the mean number of points 
$\bra N_{\ell} \ket$
and its variance $\sigma^2_{\ell}$. 
In a hyperuniform system, as 
the window size $\ell \to \infty$, 
$\sigma^2_{\ell}$ grows more slowly than 
$\bra N_{\ell} \ket$
which is proportional to the window volume.
Typical disordered systems (e.g., liquids and structured glasses)
have the standard scaling behavior 
$\sigma^2_{\ell} \simeq \ell^d$
as $\ell \to \infty$, and hence they are not hyperuniform.
Periodic point configuration 
(e.g., lattice points of perfect crystals)
are obviously hyperuniform systems, since
the number fluctuations are concentrated near the window boundary
and hence have the surface-area scaling 
$\sigma^2_{\ell} \simeq \ell^{d-1}$
as $\ell \to \infty$. We are interested in the hyperuniform point processes,
which are not perfect lattice, but are randomly 
distributed following some probability laws.
A variety of examples of disordered or random point processes
showing hyperuniformity have been reported in the review paper
by Torquato \cite{Tor18}.
A typical example on the two-dimensional plane,
which has been well studied in mathematical physics, is the complex
\textit{Ginibre point process} (GPP) \cite{Gin65}.
This is realized as the statistical ensemble of
eigenvalues of non-Hermitian random matrices 
on the complex plane $\C$, which can be
identified with the two-dimensional space $\R^2$.
First we consider an $n \times n$ matrix
whose entries are all independently 
and identically distributed (i.i.d.) 
complex standard Gaussian random variables.
We have a nearly uniform distribution of eigenvalues in a disk centered at the origin with
radius proportional to $\sqrt{n}$ on $\C$.
If we consider the $n \to \infty$ limit, we have a 
translation-invariant point process 
on $\C$ \cite{Gin65}. 
It was proved that the obtained GPP is hyperuniform
with $\sigma^2_{\ell} \simeq \ell$ \cite{Shi06}.
The origin of such hyperuniformity in a random point configuration
is long-ranged repulsive interaction between any pair of points.
As a matter of fact, GPP is a determinantal point process (DPP) \cite{HKPV09},
for which we can show that a logarithmic repulsive potential 
acts between any pair of points \cite{For10,MKS21}.
It should be remarked that DPP is also called
\textit{fermion point process} \cite{ST03},
since the determinantal structure of correlation functions
is essentially equivalent with the Slater determinant
describing wave function of fermionic particle systems.
We notice that GPP has been studied as a typical example
of strongly correlated point process in two dimensions
in non-Hermitian physics \cite{AGU20},
spectral theory \cite{TE05}, random matrix theory \cite{BF24},
and so on, 
and used in many applications 
(see, for instance, the references \cite{KT12,MS14,MS16,KK21}).

\begin{figure}[ht]
\begin{center}
\includegraphics[width=0.8\textwidth]{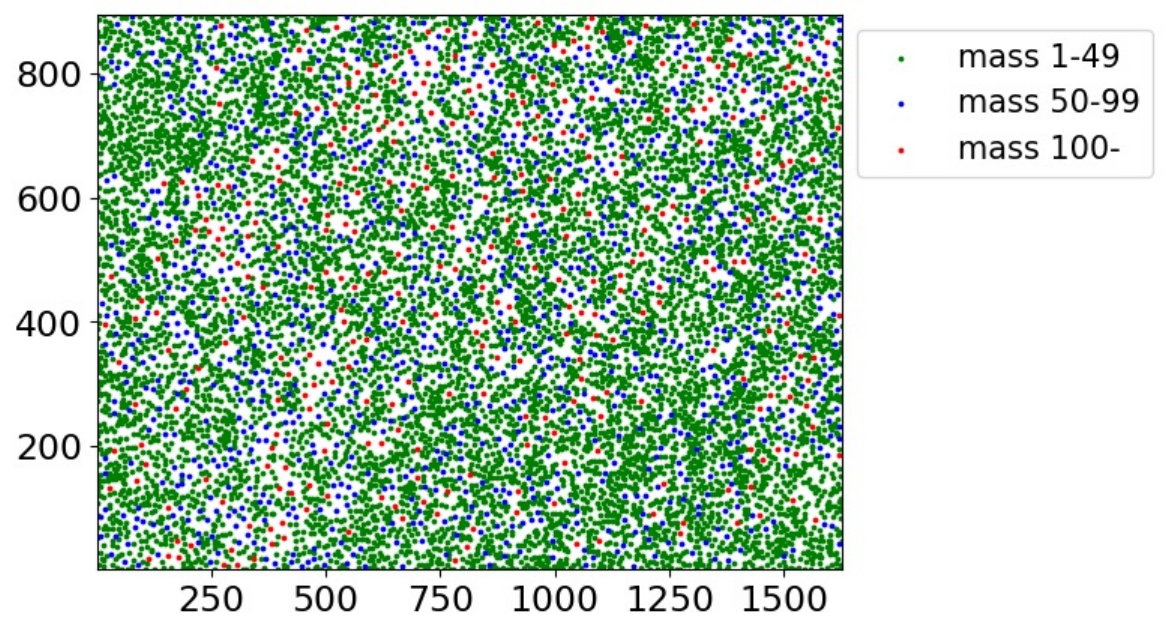}
\end{center}
\vskip -0.5cm
\caption{
A sample of point configuration of
center-of-masses of bushes in a desert.
The sizes of area are indicated not by
real lengths in meters but by numbers of pixels for
the digital data; $L_x=1626$, $L_y=895$.
The total number of points is $N=9853$ and
the density is $\rho \fallingdotseq 6.77 \times 10^{-3}$.
The bush-size (mass) distribution is represented
by colors of each points;
the points with masses $1 \leq M(\X_j) \leq 49$ are green,
the points with $50 \leq M(\X_j) \leq 99$ are blue, 
and the points with $M(\X_j) \geq 100$ are red.
See Section \ref{sec:data_analysis} showing how to 
produce such digital data.
}
\label{fig:markedPP}
\end{figure}

In the present paper, we study the spatial distributions of bushes in deserts.
In infertile lands plants can survive only if they have continued
to beat the competitions to obtain sufficient amount of
water and nutrition.
It is expected that such survival competitions
will cause strong and long-ranged repulsive interactions
among plants.
Figure~\ref{fig:markedPP} shows one sample of
point process obtained from the real data of bushes
distributed in the 2195.1 m $\times$ 1208.25 m area $\Lambda$
in a desert found in 
the Talampaya Natural Park in Argentina.
There we plotted $N=9853$ points
$\X_j \in \Lambda$, $j=1,2, \dots, N$, each of which
gives the center-of-mass of a bush.
As explained in detail in Section \ref{sec:data_analysis},
we will define a size for each bush, which we call
a \textit{mass} and write as $M(\X_j)$, $j=1,2, \dots, N$
in the present paper.
We have obtained several real samples of
point configurations of center-of-masses of
bushes $\cP=\{\X_j\}_{j=1}^{N}$ weighted
by masses $\{M(\X_j): \X_j \in \cP$\}
in deserts.
Such weighted point configurations are called
\textit{marked point processes} in 
mathematical literature \cite{DVJ03,BB10}.
The partial information of masses is shown by
coloring of points in Fig.~\ref{fig:markedPP}.

We can see in Fig.~\ref{fig:markedPP} that
the point configuration tends to sparse around the
points whose masses $M(\X_j) \geq 100$ (the bushes with large sizes
indicated by red points).
On the other hand, the points with small masses $M(\X_j) \leq 49$
(the bushes with small sizes indicated by green points)
tend to be dense and make clusters.
Therefore, if we ignore the mass data and
only look at the point configuration of 
center-of-masses of bushes, the density fluctuation
seems to be large.
Such a situation is similar to PPP and we can not
expect hyperuniformity.
In the present paper, however, we show that
if we take into account the mass data correctly,
the sample of marked point process shown 
by Fig.~\ref{fig:markedPP} as well as some other samples
obtained from the real data of bushes in deserts
are in hyperuniform states at least in the observed scales.

As briefly explained in Section \ref{sec:mathematical},
there are some families of random \textit{unmarked} 
point processes
for which hyperuniformity is mathematically proved.
To the knowledge of the present authors, however,
there is no example of random marked point process
such that it is proved to become hyperuniform
if and only if the marks to points are taken into account.
Moreover, random unmarked point processes
having hyperuniformity, which have been well studied
in mathematical physics, are all in Class I or Class II
according to the classification by Torquato \cite{Tor18}
(see Section \ref{sec:three} below).
Our data analysis of marked point processes
of bushes in deserts shows that
if they are in hyperuniform states, then
they seem to be in Class III.

In the present paper, we propose 
statistical-mechanics models in non-equilibrium
to generate random point configurations
weighted by masses on a plane,
which do not have hyperuniformity as unmarked point processes,
but have hyperuniformity as marked point processes.
In our models, we consider a stochastic processes
to iterate random \textit{thinning of points} 
\cite{Mat60,Mat86,TBB13}
and coalescing (aggregation, clustering) 
of masses \cite{Sch67,NNM95,HSIMT18}.
Our numerical simulations show that
even if we start from PPP, 
a sufficiently large number of iterations of
these two procedures generates
non-trivial marked point processes
having the desired properties.
Comparisons with the real data of bushes are discussed.

The paper is organized as follows.
In Section \ref{sec:mathematical}
we introduce mathematical expressions of
random unmarked and marked point processes
and give brief explanations of hyperuniformity.
In Section \ref{sec:data_analysis}
we explain how to produce digital data of
weighted point configurations of bushes in deserts
from the satellite images of the Google Maps.
Then we introduce a procedure to measure
hyperuniformity of the obtained data of finite sizes.
There we report the results of our data analysis for
hyperuniformity and mass distributions of
several samples obtained from plural deserts.
Section \ref{sec:models} is devoted to 
introducing our models to generate 
random marked point processes
and to showing numerical analysis of 
universal property of hyperuniformity
and non-universal property of mass distributions.
We give discussions and concluding remarks in
Section \ref{sec:discussion}.

\section{Mathematical Expressions}
\label{sec:mathematical}
\subsection{Marked and unmarked point processes}
\label{sec:pp}

Consider a continuous space $S$; for example,
the two-dimensional Euclidean space $S=\R^2$,
which can be identified with the complex plane $S=\C$.
We consider a set 
$\cP=\{\X_j \}_{j \geq 1}$ consisting of
points $\X_j \in S$, $j=1, 2, \dots$, to represent
a point configuration.
For any point $\x \in S$, we introduce a Dirac measure
(point mass) denoted by $\delta_{\x}$ \cite{DVJ03,BB10}.
This is a function of any subset $B \subset S$ such that
\begin{equation}
\delta_{\x}(B)=\begin{cases}
1, & \quad \mbox{if $\x \in B$},\\
0, & \quad \mbox{otherwise}.
\end{cases}
\label{eq:Dirac}
\end{equation}
It is useful to represent the point configuration $\cP$
by a sum of the Dirac measures as
\begin{equation}
\Xi(B)=\sum_{\X: \X \in \cP} \delta_{\X}(B),
\label{eq:Xi1}
\end{equation}
since \eqref{eq:Xi1} gives a total number of points
included in $B$ for any subset $B \subset S$;
$\Xi(B)=\sum_{\X: \X \in \cP \cap B} 1$.
We call $\Xi$ associated with $\cP$ an unmarked point process,
or simply a point process.
We also consider a marked point process, in which
each point $\X \in \cP$ carries a variable $M(\X)$.
Such a marked point process is denoted by
\begin{equation}
\Pi(B)=\sum_{\X: \X \in \cP} M(\X) \delta_{\X}(B),
\quad B \subset S.
\label{eq:Pi1}
\end{equation}
In this paper, we call $M(\X)$ a \textit{mass} 
of the point $\X \in \cP$.
For each subset $B \subset S$, \eqref{eq:Pi1} gives a total mass
of the points included in $B$;
$\Pi(B)=\sum_{\X: \X \in \cP \cap B} M(\X)$.
From a marked point process $\Pi$, we can obtain 
a unmarked point process by deleting the information
of masses of points. We denote the unmarked point process
obtained from \eqref{eq:Pi1} as $\Xi_{\Pi}$ in this paper.

We consider two kinds of random unmarked point processes, 
the Poisson point process (PPP) and the Ginibre 
point process (GPP), 
which are well studied in probability theory and random matrix theory.
We write samples of these random point processes as
$\Xi^{\rm PPP}$ and $\Xi^{\rm GPP}$, respectively.

\vskip 0.3cm
\noindent{\bf Remark 1} \,
If we consider a pair of $\X_j \in S$
and $M(\X_j) \in \R$ as a point
$\widehat{\X}_j$ in the direct product space
$S \times \R$;
$(\X_j, M(\X_j)) \Leftrightarrow \widehat{\X}_j$, 
then \eqref{eq:Pi1} can be regarded
as a usual point process in $S \times \R$
and written as 
$\Pi = \sum_{j} \delta_{\widehat{\X}_j}$ \cite{BB10}.
In the present paper, however, 
we call the point process which can be written 
in the form \eqref{eq:Pi1}
a \textit{marked point process}
and the point process in the simpler form \eqref{eq:Xi1}
as an \textit{unmarked point process}
in order to clearly distinguish these two 
types of point processes.

\subsection{Three classes of hyperuniformity}
\label{sec:three}

Let $\Lambda_{\ell}$ be a subset of $S$ with a linear size $\ell >0$. 
Such a subset of $S$ is called a \textit{window} in the
study of hyperuniformity. 
The expectation of the number of unmarked points 
of $\Xi$ included in the window $\Lambda_{\ell}$
is written as $\bra \Xi(\Lambda_{\ell}) \ket$ and its variance is 
given by
\begin{equation}
\Var[\Xi(\Lambda_{\ell})]
=\bra(\Xi(\Lambda_{\ell})-\bra \Xi(\Lambda_{\ell}) \ket)^2 \ket
\label{eq:var1}
\end{equation}
Similarly, the expectation of the total mass of
the marked points of $\Pi$ included in 
the window $\Lambda_{\ell}$ 
is written as
$\bra \Pi(\Lambda_{\ell}) \ket$ and its variance is 
given by
\begin{equation}
\Var[\Pi(\Lambda_{\ell})]
=\bra(\Pi(\Lambda_{\ell})-\bra \Pi(\Lambda_{\ell}) \ket)^2 \ket. 
\label{eq:var2}
\end{equation}
We consider the ratios,
\begin{equation}
R_\ell^\mathrm{point} = \frac{\Var [\Xi(\Lambda_\ell)]}{\bra \Xi(\Lambda_\ell) \ket}, \quad
R_\ell^\mathrm{mass} = \frac{\Var [\Pi(\Lambda_\ell)]}{\bra \Pi(\Lambda_\ell) \ket}.
\label{eq:R1}
\end{equation}
We assume that 
the window will cover the whole space
asymptotically, $\Lambda_\ell \to S$, as 
the linear size of window $\ell \to \infty$.
If 
\begin{align}
    \lim_{\ell \to \infty} R_\ell^\sharp = 0, \qquad \mbox{$\sharp=$ mass or point}, 
    \notag
\end{align}
then the point process is said to be hyperuniform.
Torquato proposed three classes depending on the order of convergence to $0$ \cite{Tor18}.
\begin{description}
    \item[Class I:] $R_\ell^\sharp \simeq \ell^{-1}$ 
    as $\ell \to \infty$. 
\vskip 0.3cm
Example 1: 
GPP $\Xi^{\rm GPP}$ \cite{Tor18,Shi06,OS08,Shi15},
\quad $S = \R^{2} \simeq \C, 
\, \, \mbox{$\sharp=$ point}$. 

\noindent Example 2:
Heisenberg family of determinantal point processes
(DPPs) \cite{MKS21}
\[
S = \R^{2D} \simeq \C^D, 
\, D \in \N \equiv \left\{1,2,...\right\}, 
\, \, \mbox{$\sharp=$ point}.
\]
\item[Class II:]
    $R_\ell^\sharp \simeq \ell^{-1} \log \ell$ 
    as $\ell \to \infty$. 
\vskip 0.3cm
\noindent Example 3: 
sinc (sine) DPP 
\cite{CL95,Meh04,ST03,Sosh00,Sosh02},
\quad $S = \R, \, \, \mbox{$\sharp=$ point}$.

\noindent Example 4: 
Euclidean family of DPPs \cite{KLS24},
\quad $S = \R^d, \, d \in \N, 
\, \, \mbox{$\sharp=$ point}$.

\item[Class III:] $R_\ell^\sharp \simeq \ell^{-\alpha}, \ 0<\alpha<1$ \ as $\ell \to \infty$.
\end{description}
\section{Data Production and Numerical Analysis 
of Point Processes}
\label{sec:data_analysis}
\subsection{Bush distributions in deserts}
\label{sec:preparation}
First we explain how to produce digital data of point configurations of bushes in a desert. 
From now on, the size of the digital data of the point configuration is written as $L_x \times L_y$. 
and the total number of points is $N$. 
We define the aspect ratio as $\lambda=L_y/L_x$. 

\begin{figure}[ht]
\begin{center}
\includegraphics[width=0.9\textwidth]{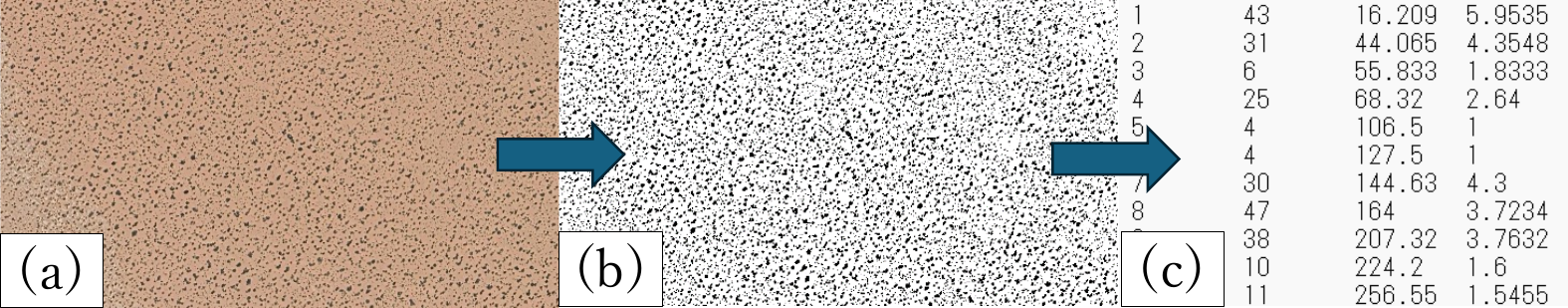}
\end{center}
\caption{
Image processing.
The Google Maps satellite image (a)
was converted to 
a gray-scale image (b) by the software \textit{Image J},
and then the digital data (c) was produced by
\textit{Matlab}.}
\label{fig:preparation_processes}
\end{figure}

The center-of-mass coordinates and sizes were obtained for bushes in deserts by the following procedure:
Here we explain the procedure using the example of a desert in the Talampaya Natural Park in Argentina.
\begin{description}
\item{(i)} \, 
A rectangular area was cut out from the Google Maps satellite image of a part of desert. In this example, the rectangular area is an image of a part of the desert centered at the point ($29^\circ 56' 02.2'' \text{S}, 67^\circ 50' 13.6'' \text{W}$) with length of $2195.1$ m in the east-west (EW) direction and length of $1208.25$ m in the north-south (NS) direction.
    Bushes are dotted with a variety of sizes as shown 
in Fig.\ref{fig:preparation_processes} (a).
\item{(ii)} \,
    The image was converted to be gray-scaled using the average method in the computer software \textit{Image J}. 
Figure \ref{fig:preparation_processes} (b) is the result obtained from Fig.\ref{fig:preparation_processes} (a). The resolution of this image is $L_x=1626$ pixels by $L_y=895$ pixels, which means that each individual pixel corresponds to a geographical area of about $1.35$m $\times$ $1.35$m.
\item{(iii)} \,
    The obtained gray-scaled image was then binarized by automatically setting the threshold value and using the Otsu method, also in \textit{Image J}. Each pixel is binarized as follows:
    \begin{align}
    \mbox{dark pixel (bush)} &\longrightarrow 
    \mbox{black pixel (mass $=1$)},
    \nonumber\\
    \mbox{light pixel (non-bush)} &\longrightarrow 
    \mbox{white pixel (mass $=0$)}.
    \nonumber
    \end{align}
\item{(iv)} \,
    The number of black pixels forming a cluster 
    (connected component) 
    was counted as a mass of each bush.
By averaging of coordinates of the black pixels 
included in a cluster, the $x$- and $y$-coordinates of the 
center-of-mass were calculated for each bush.
This procedure was done 
by \textit{Matlab} image recognition.
Figure \ref{fig:preparation_processes} (c) shows
a part of the obtained table 
which lists out the identification numbers of the bushes, 
the masses, the \textit{x}-coordinates, and the \textit{y}-coordinates of the center-of-masses.
\end{description}

Figure \ref{fig:markedPP} in Section 
\ref{sec:introduction} demonstrated the obtained
digital data of configuration of bushes in
a desert in the Talampaya Natural Park in Argentina.
The sizes of the area are given in pixel unit as
$L_x=1626$ and $L_y=895$.
There $N=9853$ dots are plotted, each of which
indicates the location of center-of-mass of a bush.
Each dot $\X_j$, $j=1,2, \dots, N$, has a data of mass 
$M(\X_j) \in \{1,2, \dots\}$.
In Fig,~\ref{fig:markedPP}, however, 
only partial information of masses
is shown by coloring of dots by
green for $1 \leq M(\X_j) \leq 49$,
by blue for $50 \leq M(\X_j) \leq 99$,
and by red for $M(\X_j) \geq 100$.

\begin{table}[hbtp]
  \caption{Data of samples}
  \label{table:desert_data}
  \centering
  \begin{footnotesize}
  \begin{tabular}{|l|l|rr|rr|rr|r|}
    \hline
    No. & \multicolumn{1}{c}{name} 
    & \multicolumn{2}{|c|}{location of center}
    & \multicolumn{2}{|c|}{size in meters}
    & \multicolumn{2}{|c|}{size in pixels} 
    &\multicolumn{1}{c|}{$N$} 
    \\ 
    \cline{3-8}
      &  & latitude & longitude & EW & NS 
    & $L_x$ & $L_y$ & \\
    \hline \hline
    1  &  (Argentina) bush 1 & 
    $29^\circ56'02.2''$S & $67^{\circ}50'13.6''$W 
    & 2195 & 1208
    & 1626 & 895 & 9853 \\
    \hline
    2  &  (Argentina) bush 2 & 
    $29^\circ55'51.5''$S & $67^{\circ}50'13.6''$W 
    & 2160 & 1080
    & 1600 & 800 & 9083 \\
    \hline
    3  &  (Argentina) bush 3 & 
    $29^\circ56'02.0''$S & $67^{\circ}49'50.1''$W 
    & 2160 & 1080
    & 1600 & 800 & 10896 \\
    \hline
    4  &  (Argentina) bush 4 & 
    $29^\circ56'12.3''$S & $67^{\circ}50'13.5''$W 
    & 2160 & 1080
    & 1600 & 800 & 10034 \\
    \hline
    5  &  (Argentina) bush 5 & 
    $29^\circ56'02.6''$S & $67^{\circ}50'39.5''$W 
    & 2160 & 1080
    & 1600 & 800 & 10249 \\
    \hline
    \hline
    6  &  Algeria 1 & 
    $28^\circ48'00.3''$N & $6^{\circ}23'42.5''$W 
    & 1786 & 198
    & 1323 & 147 & 948 \\
    \hline
    7  &  Algeria 2 & 
    $28^\circ48'15.1''$N & $6^{\circ}24'12.5''$W 
    & 201 & 2165
    & 149 & 1604 & 816 \\
    \hline
    8 &  Australia & 
    $14^\circ34'41.4''$S & $133^{\circ}31'41.1''$E
    & 2160 & 1080
    & 1600 & 800 & 3597 \\
    \hline
    9 &  Kenya & 
    $3^\circ31'00.1''$N & $38^{\circ}12'24.3''$E 
    & 2160 & 1080
    & 1600 & 800 & 10660 \\
    \hline
  \end{tabular}
  \end{footnotesize}
\end{table}

We will call the sample shown by Fig,~\ref{fig:markedPP}
simply (Argentina) \textit{bush 1} in the following.
By the similar procedure explained above,
we have prepared other four samples from the same
desert in Argentina, 
two samples from a desert in Algeria,
one from Australia, and one from Kenya.
The detailed data are listed out in 
Table \ref{table:desert_data}: 
The areas of Nos.1--5 are in 
the Talampaya Natural Park, La Rioja, Argentina, 
Nos.5 and 6 in 
the Parc culturel national de Tindouf, Oum El Assel, Algeria,
No. 7 at 
Flying Fox, Northrm Territory, Australia,
and No. 8 at Maikona, Kenya. 
The latitude and longitude are for the center of each area.
Real sizes in the east-west (EW), and north-south (NS)  directions
are given in the metric unit, whose pixels in the digital data
are also shown. The total number of bushes in each area
is given by $N$.
The areas of \textit{Algeria 1, 2} are found in
a Wadi and hence the aspect ratios 
$\lambda=L_y/L_x$ are quite deviated
 from 1.

\subsection{Poisson point process and Ginibre point process}
\label{sec:PPP_GPP}
\begin{figure}[ht]
    \begin{tabular}{cc}
      \begin{minipage}[t]{0.4\hsize}
        \centering
        \includegraphics[keepaspectratio, scale=0.4]{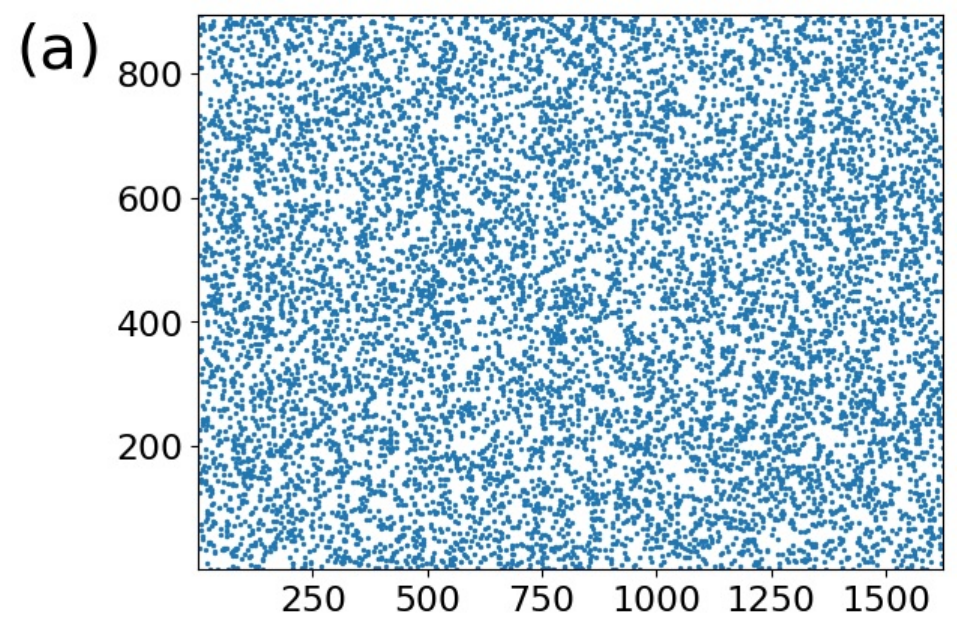}
      \end{minipage} &
      \begin{minipage}[t]{0.4\hsize}
        \centering
         \includegraphics[keepaspectratio, scale=0.4]{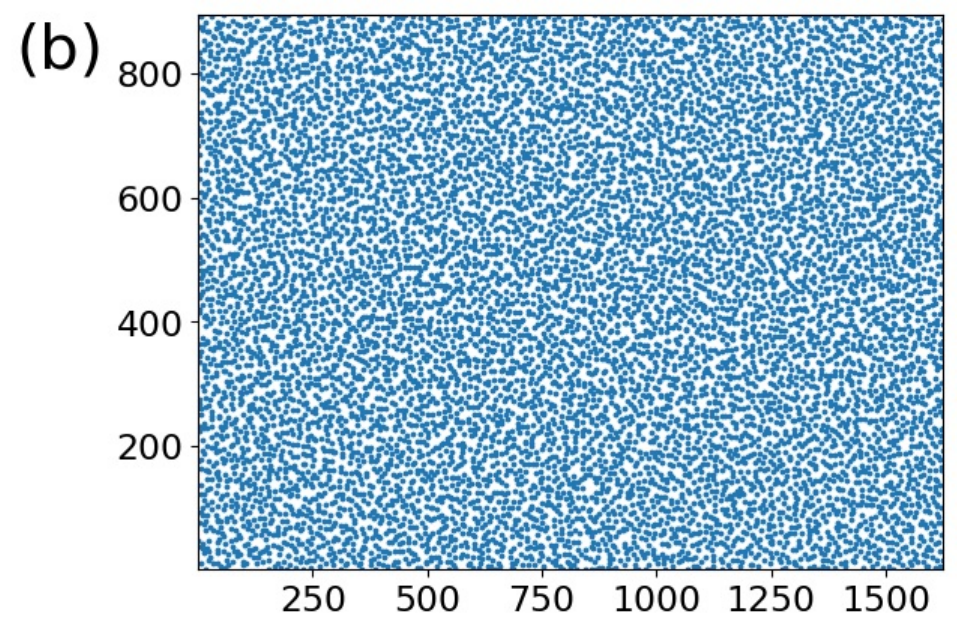}
      \end{minipage}
    \end{tabular}
    \caption{
    Numerically obtained samples of 
    (a) Poisson point process (PPP) $\Xi^{\rm PPP}$ 
    and (b) Ginibre point process (GPP) $\Xi^{\rm GPP}$.
Both are prepared in the region 
$L_x \times L_y=1626 \times 895$ with density
 $\rho \fallingdotseq 6.77 \times 10^{-3}$
 matched with the digital data for \textit{bush 1}
 shown by Fig.~\ref{fig:markedPP}.}
\label{fig:Poisson&Ginibre}
\end{figure}
In order to compare the statistics, 
here we make two digital data of
the Poisson point process (PPP)
and the Ginibre point process (GPP)
with approximately 
the same number of points $N$ and the same sizes $L_x \times L_y$ with the sample \textit{bush 1}.
Notice that they are both unmarked point processes.

PPP is a completely random configuration of points \cite{DVJ03,BB10}. 
For each point $\X_j=(X_j, Y_j)$, the coordinates
$X_j$ and $Y_j$ are chosen independently from a uniform distributions in $[0, L_x]$ and 
$[0,L_y]$, respectively. We repeat the random choosing of point $N$ times independently.
Figure \ref{fig:Poisson&Ginibre} (a) 
shows an obtained
sample for $L_x=1626$, $L_y=895$, and
$N=9853$.
The density is $\rho \fallingdotseq 6.771 \times 10^{-3}$. 
The obtained PPP is denoted by $\Xi^{\rm PPP}$.

GPP is obtained by the eigenvalue distribution of a non-Hermitian random matrix on a complex plane \cite{Gin65}. 
We consider an $n \times n$ matrix $Z_n$ as follows,
\[
Z_n = \begin{bmatrix}
\frac{1}{\sqrt{2}}(X_{11} + i Y_{11}) & \cdots 
& \frac{1}{\sqrt{2}}(X_{1n} + i Y_{1n}) \\
\vdots & \ddots & \vdots \\
\frac{1}{\sqrt{2}}(X_{n1} + i Y_{n1}) & \cdots 
& \frac{1}{\sqrt{2}}(X_{nn} + i Y_{nn})
\end{bmatrix},
\]
where $i=\sqrt{-1}$.
Here $X_{jk}$ and $Y_{jk}$, $j, k=1, 2, \dots, n$ are
i.i.d.~following the standard normal distribution $\mathrm{N}(0,1)$, 
that is, 
the Gaussian distribution with mean zero and variance $\sigma^2=1$.
It is known as the \textit{circular law} 
such that the $n$ eigenvalues of $Z_n$ are not degenerated and uniformly distributed in a disk centered at the origin with radius $\sqrt{n}$, if $n$ is sufficiently large \cite{For10, Meh04}. 
The circular law implies that the density of GPP is
\begin{align}
    \rho_\mathrm{G} 
    = \frac{n}{\pi (\sqrt{n})^2} 
    = \frac{1}{\pi}.
    \notag
\end{align}
Inside of the disk, we prepare a rectangular region $L_x^0 \times L_y^0$ with the aspect ratio $L_y^0 / L_x^0 = \lambda$ 
so that it contains $N$ eigenvalues.
These conditions determine that
\begin{align}
    \rho_\mathrm{G} L_x^0 L_y^0 
    = N \ \Longrightarrow \ L_x^0 = \sqrt{\frac{N \pi}{\lambda}}.
    \label{eq:rho_G}
\end{align}
If we put $N=9853$ and
$\lambda=L_y/L_x=895/1626$,  
\eqref{eq:rho_G} gives
\begin{align}
    L_x^0 
    = \sqrt{\frac{N \pi}{\lambda}} 
    = \sqrt{N \pi \frac{L_x}{L_y}} \fallingdotseq 237.1.
    \notag
\end{align}
In our numerical calculation, we have used $Z_n$ with $n=2.2 \times 10^4$.
Hence we see $237.1 < 2 \sqrt{n} \fallingdotseq 296$,
and confirmed that the matrix size is large enough
for our purpose.
Then we dilate the obtained point configuration in the region $L_x^0 \times L_y^0$ to that in 
$L_x \times L_y$; that is, the size of region is enlarged 
by linear factor $L_x / L_x^0 \fallingdotseq 6.857$.
The obtained sample is shown by 
Fig.\ref{fig:Poisson&Ginibre} (b). 
The expected density of dots is 
$\rho = (1/\pi) \times 6.857^{-2}
\fallingdotseq 6.770 \times 10^{-3}$.
The exact number of points in 
Fig.\ref{fig:Poisson&Ginibre} (b) is 9862,
which has a small deviation from the target number
$N=9853$ due to a sample fluctuation and
the finite-size effect of matrix.
But this deviation is negligible in our scaling analysis
mentioned below.
The obtained GPP is denoted by $\Xi^{\rm GPP}$.

We can see clear difference between PPP and GPP
in Fig.~\ref{fig:Poisson&Ginibre}.
Sparseness found in PPP is much suppressed in GPP.
As already mentioned in previous sections,
GPP is hyperuniform, while PPP is not.
If we ignore the coloring of points in Fig.~\ref{fig:markedPP}
for \textit{bush 1},
the point configuration seems to be similar to PPP.
The problem is how about the point configuration
if we take into account the information of masses of 
bushes. Is it possible to be in hyperuniform state
as a marked point process?
The answer is positive as explained below.

\subsection{Measurements of hyperuniformity 
by finite-size data}
\label{sec:measurements}
We prepare a series of windows
parameterized by linear size $\ell$, 
$\{\Lambda_{\ell}\}$,  such that $\Lambda_{\ell}$ 
expands monotonically as $\ell$ increases and
$\Lambda_{\ell}$ will cover the whole space
asymptotically in the limit $\ell \to \infty$.
As explained in Section \ref{sec:three}, 
the hyperuniformity of random point process
is defined and classified mathematically by 
asymptotics in the limit $\ell \to \infty$ 
of the ratio $R_{\ell}$
of the variance with respect to the expectation 
concerning the 
number of points included in $\Lambda_{\ell}$.
In real experiments of physics as well as
observation in ecological systems, however, 
we have only \textit{a few number} of sample configurations
in \textit{finite regions}.

On the other hand, two samples
of point processes given by Fig.~\ref{fig:Poisson&Ginibre}
can be distinguished from each other,
since they represent typical properties of
uncorrelated and fluctuating point configuration
in space in Fig.~\ref{fig:Poisson&Ginibre} (a)
and of correlated point configuration where
fluctuation is suppressed to keep hyperuniformity
in Fig.~\ref{fig:Poisson&Ginibre} (b).
By such observations, here we propose a numerical
method to measure hyperuniformity 
using only one but typical sample of
point processes observed in nature.
We adopted two strategies;
(i) Instead of observing the asymptotics 
in large-scale limit, we divide a given finite system
into smaller subsystems systematically.
(ii) Instead of averaging over many samples,
we average the data over subsystems obtained
by division of one sample.

\begin{figure}[ht]
\begin{center}
\includegraphics[width=0.6\textwidth]{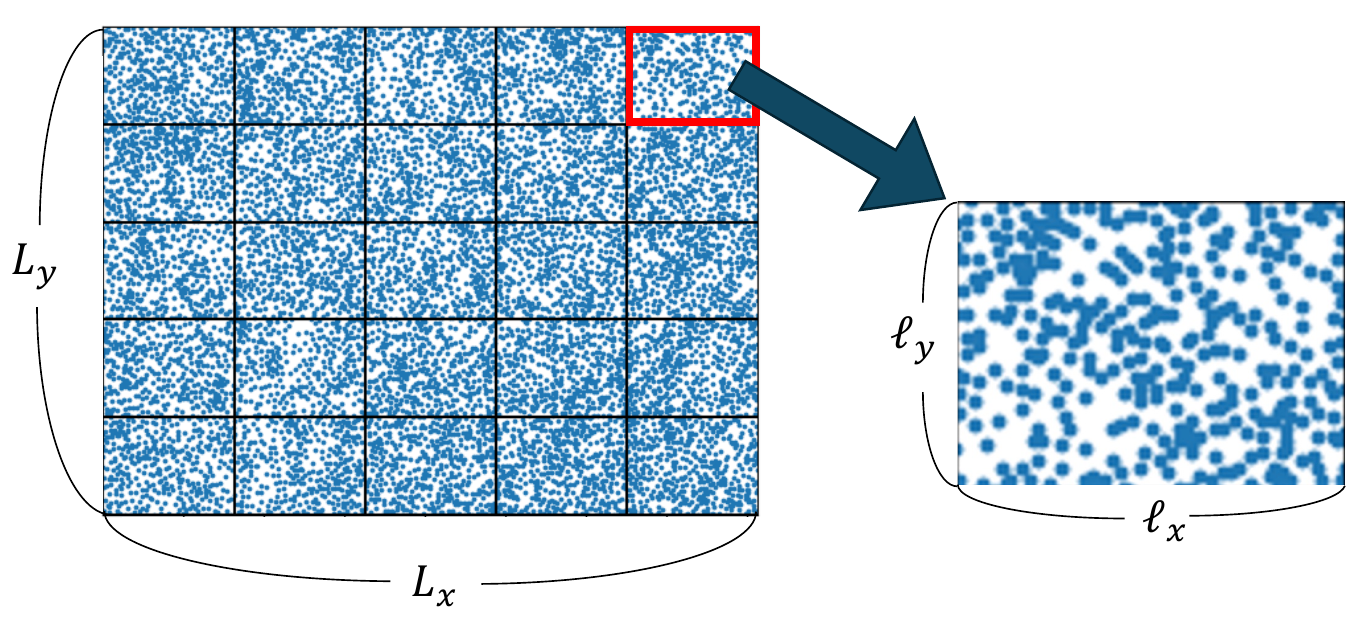}
\end{center}
\caption{
The region $L_x \times L_y$ 
on which a sample of point process is given is divided into
$m \times m$ subregions.
Each subregion is a similar rectangular 
$\ell_x \times \ell_y$ denoted by
$\Lambda_{\ell_x}^{j}$, $j=1,2, \dots, m^2$. 
The case with $m=5$ is shown.}
\label{fig:devision_image}
\end{figure}

We consider a rectangular region $L_x \times L_y$. 
We fix the aspect ratio $\lambda=L_y/L_x$ and 
write it simply as $\Lambda_{L_x}$
putting only the size in the $x$-direction
as a subscript.
Let $m \in \N$ and
we divide $\Lambda_{L_x}$
into $m \times m$ subregions.
The lengths in the $x$- and $y$-directions are then
given by
\[
    \ell_x = L_x/m
\quad
    \ell_y = L_y/m.
\]
See Fig.\ref{fig:devision_image}. 
The obtained subregions are labeled by
$j=1,2, \dots, m^2$ as
\begin{align}
    \Lambda_{L_x} \quad
    \Longrightarrow \quad
    \left\{ \Lambda_{\ell_x}^1, \Lambda_{\ell_x}^2, 
      \dots, \Lambda_{\ell_x}^{m^2} \right\}.
    \notag
\end{align}

First we consider an unmarked point process
$\Xi(\Lambda_{L_x})$.
Examples are the point configurations
of center-of-mass coordinates of
bushes in a desert $\Xi_{\Pi}$ 
for which the mass information is deleted, 
$\Xi^{\rm PPP}$, and $\Xi^{\rm GPP}$.
For all $j=1,2, \dots, m^2$, the expectation
of $\Xi(\Lambda_{\ell_x}^{j})$ is simply evaluated as
$\bra \Xi(\Lambda_{\ell_x}^{j}) \ket = \rho \ell_x \ell_y$,
where $\rho$ is the density of points
of the given sample $\Lambda_{L_x}$,
$\rho=\Xi(\Lambda_{L_x})/(L_x L_y)
=N/(L_x L_y)$.
Then for each subregion $\Lambda_{\ell_x}^j$,
$j=1, 2, \dots, m^2$, we count the deviation of
number of points observed in it from the expectation, 
and define the variance by the following arithmetic mean
over $m^2$ subregions,
\begin{equation}
\Var [\Xi (\Lambda_{\ell_x})] =
\frac{1}{m^2} \sum_{j=1}^{m^2} 
\left( \Xi (\Lambda_{\ell_x}^j) - \rho \ell_x \ell_y \right)^2.
\label{eq:Var1}
\end{equation}
The ratios of the variances with respect to expectations
are written as
$R^{\rm point}_{\ell_x}$ for unmarked point process of
bush configurations $\Xi_{\Pi}$,
$R^{\rm PPP}_{\ell_x}$ for $\Xi^{\rm PPP}$,
and $R^{\rm GPP}_{\ell_x}$ for $\Xi^{\rm GPP}$;
\begin{equation}
R^{\rm point}_{\ell_x}
=\frac{\Var[\Xi_{\Pi}(\Lambda_{\ell_x})]}{\rho \ell_x \ell_y},
\quad
R^{\rm PPP}_{\ell_x}
=\frac{\Var[\Xi^{\rm PPP}(\Lambda_{\ell_x})]}{\rho \ell_x \ell_y},
\quad
R^{\rm GPP}_{\ell_x}
=\frac{\Var[\Xi^{\rm GPP}(\Lambda_{\ell_x})]}{\rho \ell_x \ell_y}.
\label{eq:R_point}
\end{equation}

Next we consider a marked point process
$\Pi(\Lambda_{L_x})$.
Examples are the point configurations
of center-of-mass coordinates of
bushes in a desert $\Xi_{\Pi}$ 
weighted by masses.
Instead of the density of point $\rho$, 
here we calculate the \textit{density of mass}
defined by
\[
\rho^{\rm mass} = 
\frac{\Pi(\Lambda_{L_x})}{L_x L_y}
=\frac{\sum_{\X: \X \in \cP \cap \Lambda_{L_x}} M(\X)}
{L_x L_y}.
\]
For all $j=1,2, \dots, m^2$, the expectation
of $\Pi(\Lambda_{\ell_x}^{j})$ is evaluated as
$\bra \Pi(\Lambda_{\ell_x}^{j}) \ket = \rho^{\rm mass} \ell_x \ell_y$.
Then the variance is calculated as
\begin{equation}
\Var [\Pi(\Lambda_{\ell_x})] =
\frac{1}{m^2} \sum_{j=1}^{m^2} 
\left(\Pi(\Lambda_{\ell_x}^j)) 
- \rho^{\rm mass} \ell_x \ell_y \right)^2.
\label{eq:Var2}
\end{equation}
The ratio of the variance with respect to expectation
is written as
$R^{\rm mass}_{\ell_x}$ for 
bush configurations weighted by masses;
\begin{equation}
R^{\rm mass}_{\ell_x}
=\frac{\Var[\Pi(\Lambda_{\ell_x})]}{\rho^{\rm mass} \ell_x \ell_y}.
\label{eq:R_mass}
\end{equation}

\subsection{Hyperuniformity of 
marked point processes
of bushes in a desert in Argentina}
\label{sec:result1}
\begin{figure}[htbp]
\begin{center}
  \begin{minipage}[b]{0.5\linewidth}
    \centering
    \includegraphics[keepaspectratio, scale=0.5]
    {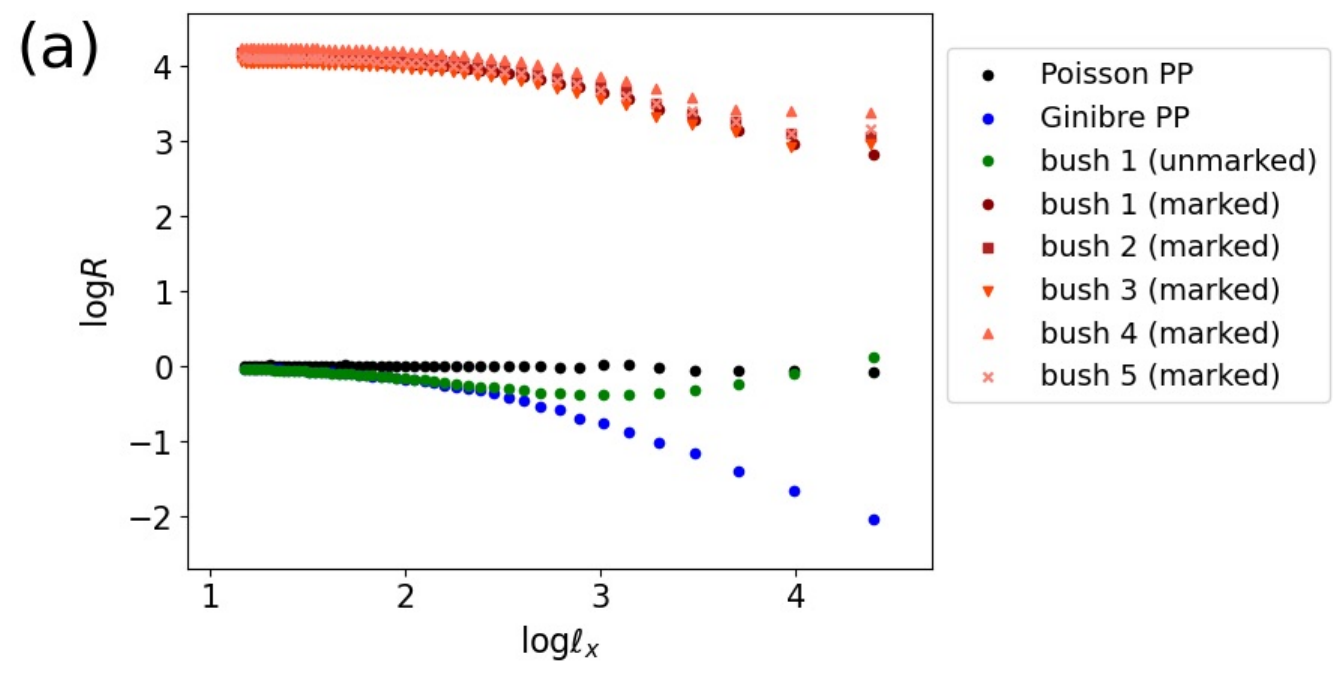}
  \end{minipage}
  \\
  \begin{minipage}[b]{0.5\linewidth}
    \centering
    \includegraphics[keepaspectratio, scale=0.5]
    {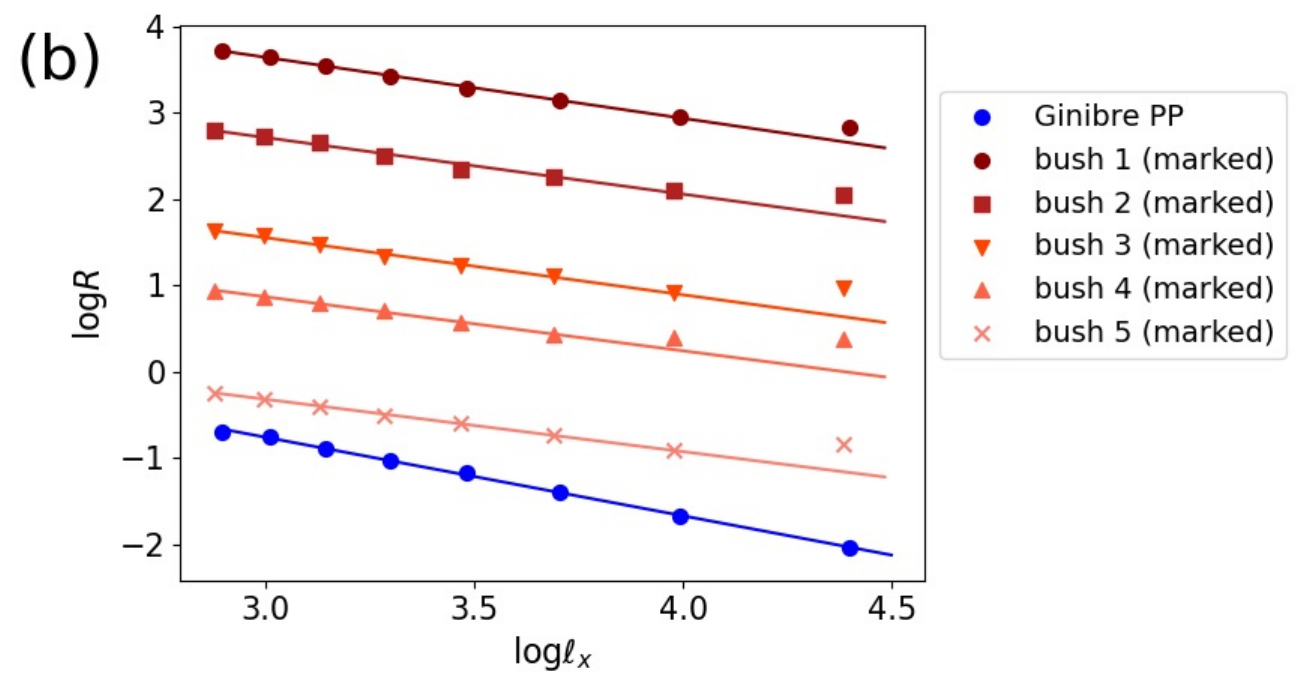}
  \end{minipage}
  \caption{
  Numerical measurements of hyperuniformity using finite-size data.
  (a) Log-log plots of $R_{\ell_x}$ versus $\ell_x$
for various unmarked and marked point processes.
The bush configuration not taking into account 
of mass information 
does not show hyperuniformity similarly to PPP, while 
the bush configurations weighted by masses 
exhibit hyperuniformity similarly to GPP.
  (b) Linear fitting to \eqref{eq:fitting} was performed 
for the interval of $X \equiv \log \ell_x$ such that
$2.8 \leq X \leq 4.4$ for GPP,
$2.8 \leq X \leq 4.0$ for the marked point processes of 
\textit{bush~1-3} and \textit{5}, and
$2.8 \leq X \leq 3.8$ for \textit{bush 4}.
The fitting works well and
the exponent $\alpha$ is evaluated as \eqref{eq:alpha}.
Here the dots are shifted 
in the direction of $Y \equiv \log R$
by $-(n-1)$ for each \textit{bush n}, $n=1,2,\dots, 5$, 
respectively, 
to avoid overlaps of plots and fitting lines.
}
\label{fig:loglog}
\end{center}
\end{figure}

In Fig.~\ref{fig:loglog} (a), we show 
dependence on $\log \ell_x$ of 
logarithms of $R^{\rm point}_{\ell_{x}}$
for the unmarked point configurations 
of the sample \textit{bush 1}, 
$\log R^{\rm PPP}_{\ell_x}$ and
$\log R^{\rm GPP}_{\ell x}$.
For the totally five samples from
the desert in Argentina, 
\textit{bush 1--5}
in Table~\ref{table:desert_data},
$\log R^{\rm mass}_{\ell_{x}}$
are also plotted versus $\log \ell_x$ 
in Fig.~\ref{fig:loglog} (a), 
which are for the marked point processes. 
In the numerical measurements of 
variances by \eqref{eq:Var1} and \eqref{eq:Var2}
in Section \ref{sec:measurements},
the numbers of subregions $m^2$ become relatively small
for large values of $\ell_x$, and hence
the obtained plots are scattered.
For this reason we only use the data with
$\log \ell_x < 4.5$.

With increment of the window size $\ell_x$, 
the number fluctuation 
$R^{\rm PPP}_{\ell_x}$
does not show any decay, 
while $R^{\rm GPP}_{\ell_x}$ shows
a systematic decay.
These plots suggest that $\Xi^{\rm PPP}$ has
no hyperuniformity, but
$\Xi^{\rm GPP}$ is in a hyperuniform state,
as expected. 
We see that the 
behavior of $\log R^{\rm point}_{\ell_x}$ 
versus $\log \ell_x$ of
unmarked point processes of \textit{bush 1}
is very similar to
that of PPP, as already observed in Fig.~\ref{fig:markedPP}.
It implies that 
if the mass is not taken into account, 
any hyperuniformity can \textit{not} be observed in bush
configurations in the desert.
If we consider the marked point processes, however, 
in which each point is weighted by mass, 
linear decay of $\log R^{\rm pass}_{\ell_x}$
with increment of $\log \ell_{x}$ is commonly observed 
as shown by the five data from the desert in Argentina,
\textit{bush 1--5} (marked) 
in Fig.~\ref{fig:loglog} (a).

Figure~\ref{fig:loglog} (b) shows
the results of the linear fitting,
\begin{equation}
\log R_{\ell_x} = -\alpha \log \ell_x + c,
\label{eq:fitting}
\end{equation}
for GPP $\Xi^{\rm GPP}$, and \textit{bush 1--5} (marked). 
Here in order to avoid overlaps of plots and fitting lines,
the data points are shifted in the $Y$-direction
by $-(n-1)$ for data, \textit{bush n} (marked), $n=1,2,\dots, 5$.
The numerically estimated values of exponent $\alpha$
are the following; 
$\alpha^{\rm GPP} = 0.91 \fallingdotseq 1$,
$\alpha^{\rm bush \, 1}=0.71$, 
$\alpha^{\rm bush \, 2}=0.66$, 
$\alpha^{\rm bush \, 3}=0.66$, 
$\alpha^{\rm bush \, 4}=0.63$, 
and
$\alpha^{\rm bush \, 5}=0.61$.
The present numerical analysis suggest that
the bush configurations weighted by masses
sampled from the desert in Argentina
are in the hyperuniform state of Class III,
\begin{equation}
R_{\ell_x}^{\rm mass} \simeq \ell_x^{-\alpha^{\rm Argentina}}
\quad
\mbox{with} \quad
\alpha^{\rm Argentina} \fallingdotseq 0.65. 
\label{eq:alpha}
\end{equation}

\subsection{Mass distributions of 
marked point processes of bushes}
\label{sec:mass_dist}

\begin{figure}[ht]
\begin{center}
\includegraphics[width=0.6\textwidth]
{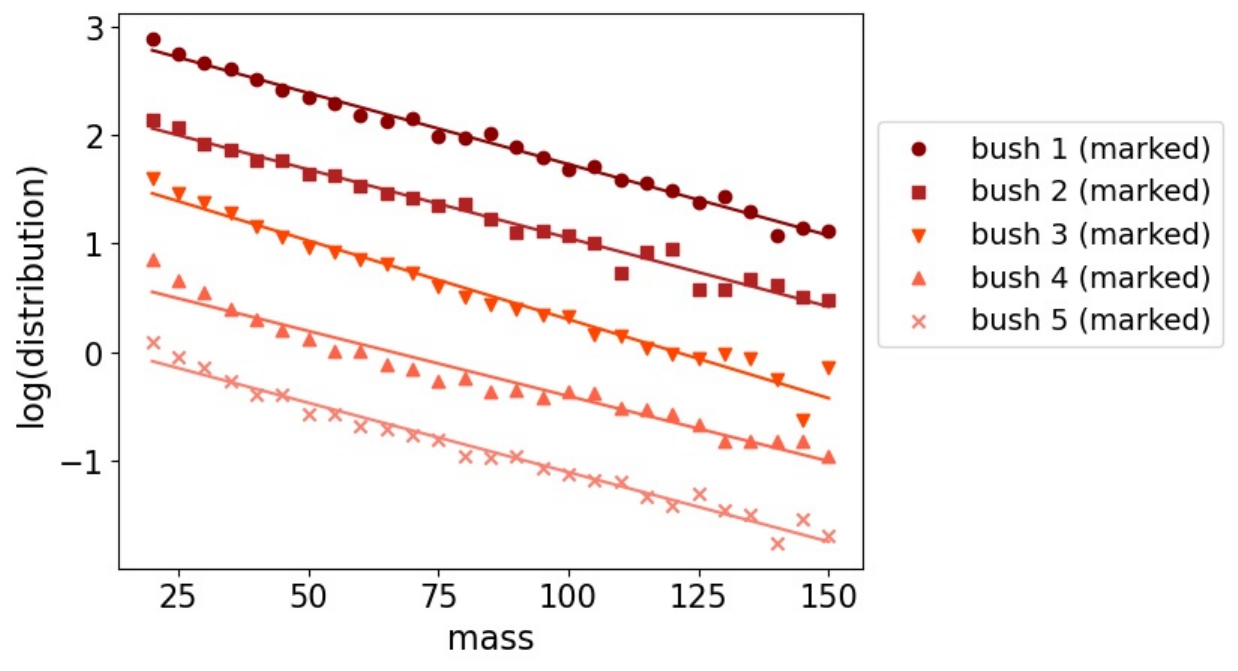}
\end{center}
\caption{
Semi-log plots of bush-mass distributions in \textit{bush 1--5}.
The dots are shifted in the direction of $\log$(distribution) 
by $-0.7 (n-1)$ for each \textit{bush n}, $n=1,2,\dots, 5$, 
respectively, 
to avoid overlaps of plots and fitting lines.
The linear fitting to \eqref{eq:Mass_dist} 
works well in the shown interval of mass, 
which gives 
$\sm_0^{\rm bush \, 1} = 76$,
$\sm_0^{\rm bush \, 2} = 79$,
$\sm_0^{\rm bush \, 3} = 69$,
$\sm_0^{\rm bush \, 4} = 83$,
$\sm_0^{\rm bush \, 5} = 78$.
} 
\label{fig:mass_dist}
\end{figure}
For the marked point processes of \textit{bush 1--5},
we have measured the mass distributions.
Figure~\ref{fig:mass_dist} shows the semi-log graphs for histograms
of $M$.
The results suggest 
that the mass distributions
are well described by the exponential distribution,
\begin{equation}
{\rm Prob}(M \in [\sm, \sm+d \sm))
\simeq e^{-\sm/\sm_0} d \sm
\quad \mbox{with} \quad
\sm_0=\sm_0^{\rm Argentina}\fallingdotseq 77.
\label{eq:Mass_dist}
\end{equation}
\subsection{Hyperuniformity of other deserts}
\label{sec:result3}
\begin{figure}[ht]
\begin{center}
\includegraphics[width=0.6\textwidth]{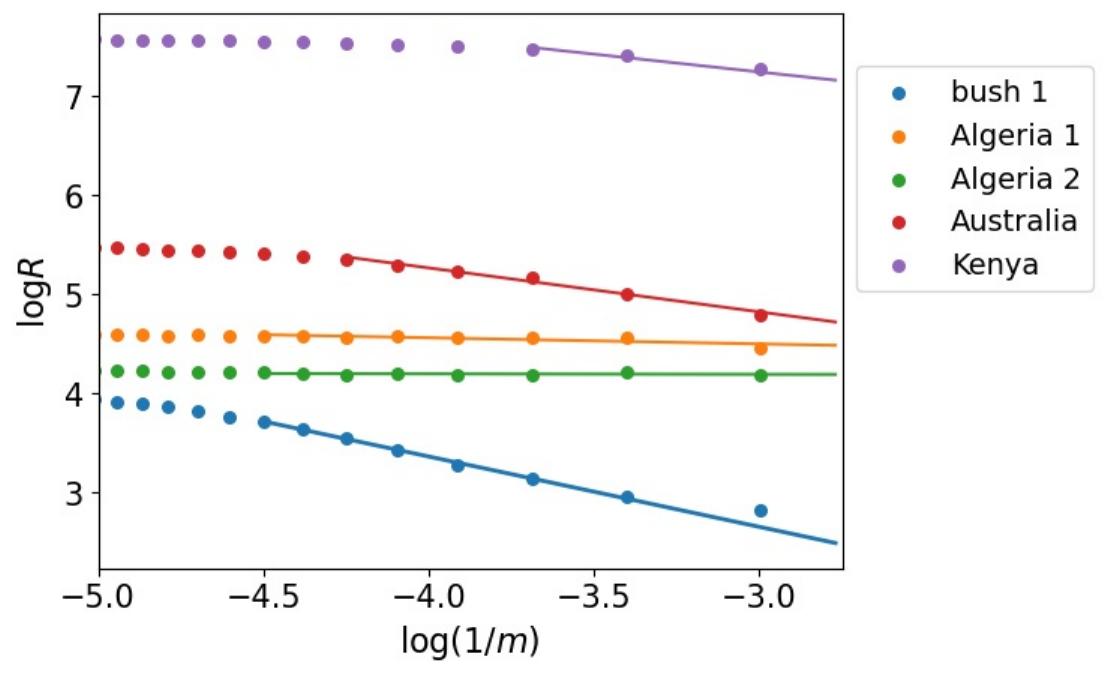}
\end{center}
\caption{
Logarithms of $R^{\rm mass}$ 
are plotted as functions of $\log (1/m)$
for a variety of samples from different deserts.
Linear fitting gives the exponent $\alpha$.
}
\label{fig:other_deserts}
\end{figure}

Figure~\ref{fig:other_deserts} shows 
logarithms of the ratios $R^{\rm mass}$ 
as functions of $\log(1/m)$ for the samples
in deserts Nos.6--9
as well as for the \textit{bush 1}
given as No.1 in Table \ref{table:desert_data}, 
where $m$ is the number of divisions 
of the finite-size data as explained in
Section \ref{sec:measurements}.
The linear fitting to 
$\log R = - \alpha \log (1/m)+c$
gives the exponents
$\alpha^{\rm Algeria \, 1} =0.062$,
$\alpha^{\rm Algeria \, 2}=0.0049$,
$\alpha^{\rm Australia}=0.44$,
and
$\alpha^{\rm Kenya}=0.36$.
The marked point processes
obtained from the desert in Algeria 
(sample Nos.6 and 7) do not show
hyperuniformity ($\alpha^{\rm Algeria} \fallingdotseq 0$),
while those from Australia (No.8) and
Kenya (No.9) seem to be in the 
hyperuniform state in Class III,
where the values of the exponent $\alpha$ 
are smaller than $\alpha^{\rm Argentina} \fallingdotseq 0.65$.

\section{Models}
\label{sec:models}
\subsection{Random thinning-coalescing processes}
\label{sec:process}
For a rectangular region with
aspect ratio $\lambda$, 
$\Lambda_{L_x} = \left\{ (x,y) \in \R^2 : 0 \leq x \leq L_x, 
0 \leq y \leq \lambda L_x \right\}$, 
we define an extended region with $r >0$ by
\begin{align}
    \Lambda_{L_x, r} = \left\{ (x,y) \in \R^2 : 
    -2r \leq x \leq L_x + 2r, -2r \leq y \leq \lambda L_x + 2r \right\} \supset \Lambda_{L_x}.
    \notag
\end{align}
For two points
$\X=(X_x, X_y)$,
$\Y=(Y_x, Y_y) \in \Lambda_{L_x, r}$, the distance is defined by
$d(\X, \Y) = \sqrt{(X_x-Y_x)^2+(X_y-Y_y)^2}$. 
Give a probability density ${p (a)}$ for a non-negative random variable ${A}$ so that
\begin{align}
    P ( A \leq a) = \int_{0}^{a} p (a') da',
    \label{eq:p}
\end{align}
and the mean is finite; 
$a_0 \equiv \bra A \ket =\int_0^{\infty} a' p(a') da'< \infty$.
Let $r_0 \equiv \sqrt{a_0/\pi}$.

We consider the following transformation $\cT$ of 
the marked point process on $\Lambda_{L_x, r_0}$, 
\begin{align*}
\cT: 
\Pi (B) = \sum_{\X:\X \in \cP} M ( \X ) \delta_{\X} (B) \quad
    \Longrightarrow \quad
\widetilde{\Pi} (B) = \sum_{\X:\X \in \widetilde{\cP}} \widetilde{M} (\X) \delta_{\X} (B),
\end{align*}
where $B$ is an arbitrary subregion of
$\Lambda_{L_x}$. 
\begin{description}
    \item{(i)} \
    Choose a point $\X$ randomly from 
    the set of points $\cP$ which are included in $\Lambda_{L_x}$; 
    $\X \in \cP \cap \Lambda_{L_x}$.
    \item{(ii)} \
    Let $A$ be a non-negative random variable following the probability law \eqref{eq:p} and set $R \equiv \sqrt{A / \pi}$.
    That is, $R$ is the radius of a disk whose area is equal to $A$.
    For a chosen point $\X$, 
    define a subset of $\cP$ by
\[
        \cP (\X) = \{ \Y \in \cP \backslash \left\{ \X \right\} : 
        d(\X, \Y) \leq R \},
\]
where $A \setminus B$ means subtraction of a set $B$
from a set $A$.
Notice that $\cP(\X)$ is a collection of all points
except $\X$, which are included in the
disk centered at $\X$ with radius $R$. 
    \item{(iii)} \
    Set 
    \begin{align}
        \widetilde{\cP} &= \cP \backslash \cP(\X), 
        \label{eq:thinning}
        \\
        \widetilde{M} (\Z) &=
    \begin{cases}
    \displaystyle{ M(\X) + \sum_{\Y:\Y \in \cP (\X)} M(\Y) }, 
    & \mbox{if ${\Z= \X}$}
    \cr
    M (\Z), & \mbox{if ${\Z \in \widetilde{\cP} 
    \backslash \left\{ \X \right\}}$}.
    \end{cases}
    \label{eq:coalescing}
    \end{align}
\end{description}
The reduction of points \eqref{eq:thinning}
represents a random \textit{thinning} \cite{Mat60,Mat86,TBB13}.
That is, all points except $\X$ included in
a disk centered at $\X$ with radius $R$
are deleted. 
Then the first line of \eqref{eq:coalescing} 
represents \textit{coalescing} (aggregation, clustering) 
of masses of the deleted points
into the chosen point $\X$ \cite{Sch67,NNM95,HSIMT18}.

As reported in the following, 
we have iterated the transformation ${\mathcal{T}}$. 
We have found that even though the initial marked point process
has no hyperuniformity, 
if we iterate the transformation $\cT$ sufficiently
many times, then we can obtain
non-trivial marked point process having hyperuniformity.
In the practical simulation, iteration of $\cT$
is performed as follows.
First we fix a rate $p \in (0, 1)$.
Assume that 
in a given marked point process 
$\Pi=\sum_{\X: \X \in \cP} M(\X) \delta_{\X}$,
the total number of points is 
$n=\sharp (\cP \cap \Lambda_{L_x})
=\Xi_{\Pi}(\Lambda_{L_x})$.
We choose $p n$ points from $\cP \cap \Lambda_{L_x}$
uniformly at random and prepare an ordered list of them
as candidates for the central point of a disk
in the procedures (ii) and (iii) in the above algorithm.
We assign the first point in the list as
$\X$ in (i) and then perform (ii) and (iii).
In the thinning \eqref{eq:thinning}, 
if some of the points in the list are deleted, then
they are deleted also in the list.
We eliminate also the used point from the list.
Next in the updated list the first point is 
assigned as $\X$ in (i) and using it as a center of a disk, 
the procedures (ii) and (iii) are performed.
By definition, the number of points in the list
decreases monotonically.
When the list becomes empty, 
we think that one iteration-term is complete and
increase the \textit{number of iteration-terms} $T$ by 
unity; $T \to T+1$.
For the updated marked point process
$\widetilde{\Pi}=
\sum_{\X: \X \in \widetilde{\cP}} 
\widetilde{M}(\X) \delta_{\X}$,
where the total number of points is now given by
$\widetilde{n}=\sharp(\widetilde{\cP} \cap \Lambda_{L_x})$,
we choose $p \widetilde{n}$ points randomly
from $\widetilde{\cP} \cap \Lambda_{L_x}$
and make a new ordered 
list of candidates for the center of a disk.
We repeat the iteration-terms of $\cT$.
By the thinning \eqref{eq:thinning},
the total number of point $n$ is decreasing,
and hence the computational time for
one iteration-term decreases
as $T$ increases, since we fix $p \in (0, 1)$.
In the following, we report the results
in the case $p=0.1$.

\subsection{Results of numerical simulations}
\label{sec:model_result1}
\begin{figure}[ht]
\begin{center}
\includegraphics[width=0.6\textwidth]
{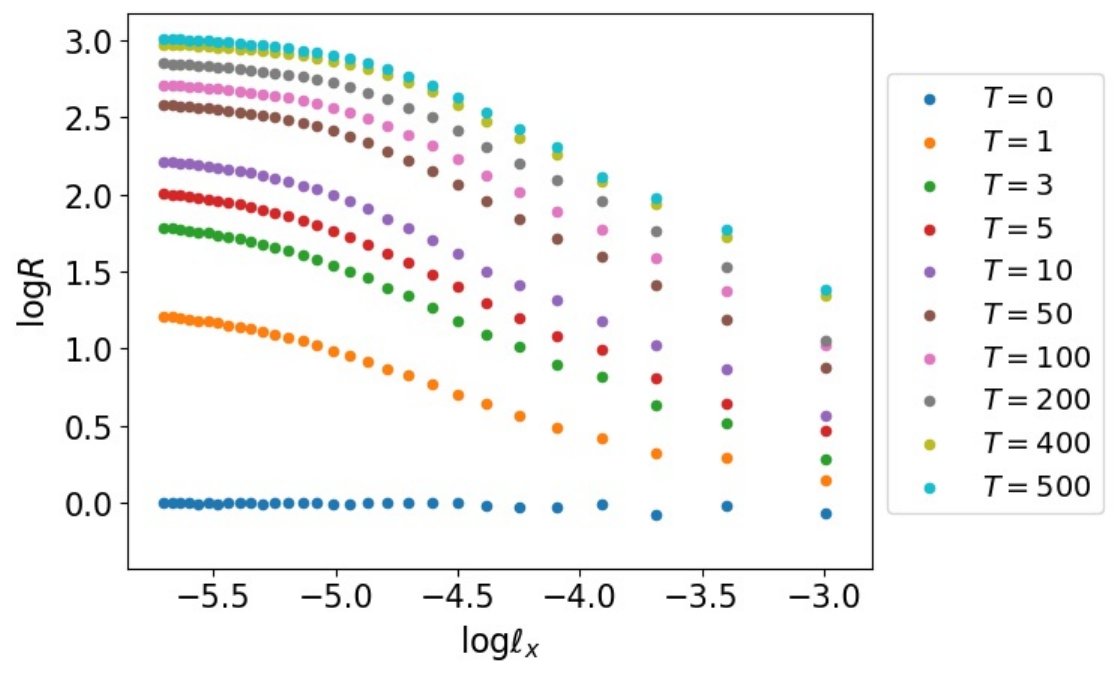}
\end{center}
\caption{
Dependence on the number of iteration-terms $T$
is shown in the $\log R_{\ell_x}$ versus $\log \ell_x$ plots,
where $L=L_x=L_y=4000$ 
and $a_0 = \theta_0=1000$ for \eqref{eq:exp_dist}.}
\label{fig:Thinning_comp}
\end{figure}

Based on the observations in
\textit{bush 1--5} reported in Section \ref{sec:mass_dist}
such that the distribution of mass $M$ is well
approximated by the exponential distribution, 
first we choose the exponential distribution for $A$,
since the mass transform $M$ 
in each coalescing \eqref{eq:coalescing}
will be given by the area $A$ multiplied by density
of points.
The probability density function in \eqref{eq:p} is assumed to be
given by
\begin{equation}
p(a)=p(a; \theta_0) = \frac{1}{\theta_0} e^{-a/\theta_0},
\label{eq:exp_dist}
\end{equation}
where $\theta_0$ is the scale parameter.
It is easy to verify that 
$a_0 \equiv \bra A \ket=\theta_0$
and $\Var[A]=\theta_0^2$.
For simplicity, in the computer simulations of our model,
we fix the aspect ratio $\lambda=1$;
that is, $L_x=L_y \equiv L$.

Let $L=4000$ and $a_0=1000$.
We start our algorithm from
a sample of PPP obtained in 
Section \ref{sec:PPP_GPP}, 
where $M(\X) = 1$ 
for all points $\X \in \cP$
at the beginning $T=0$. 
For each marked point process
obtained after $T$ iteration-terms of 
transformation $\cT$, 
we have performed the 
measurements explained in Section \ref{sec:measurements}.
The $\log R_{\ell_x}$ versus $\log \ell_x$ plots 
are given in Fig.~\ref{fig:Thinning_comp} 
from $T=0$ (the initial PPP) to $T=500$.
As $T$ increases, the value of $\log R_{\ell_x}$
increases and the region of
$\log \ell_x$ in which $\log R_{\ell_x}$ shows
a linear decay seems to extend.
Moreover, we have observed convergence of the
plots when $T \gtrsim 500$. 
We also confirmed the convergence of the plots
to the same curve 
in the simulations starting from a sample of 
GPP with $M(\X) \equiv1$ 
prepared in Section \ref{sec:PPP_GPP}.
From now on, we will report the results of
numerical simulations, which were obtained 
after $T=500$ iteration-terms
of transformation $\cT$.

\subsubsection{Hyperuniformity}
\label{sec:log_log}
\begin{figure}[htbp]
\begin{center}
  \begin{minipage}[b]{0.5\linewidth}
    \centering
    \includegraphics[keepaspectratio, scale=0.5]    {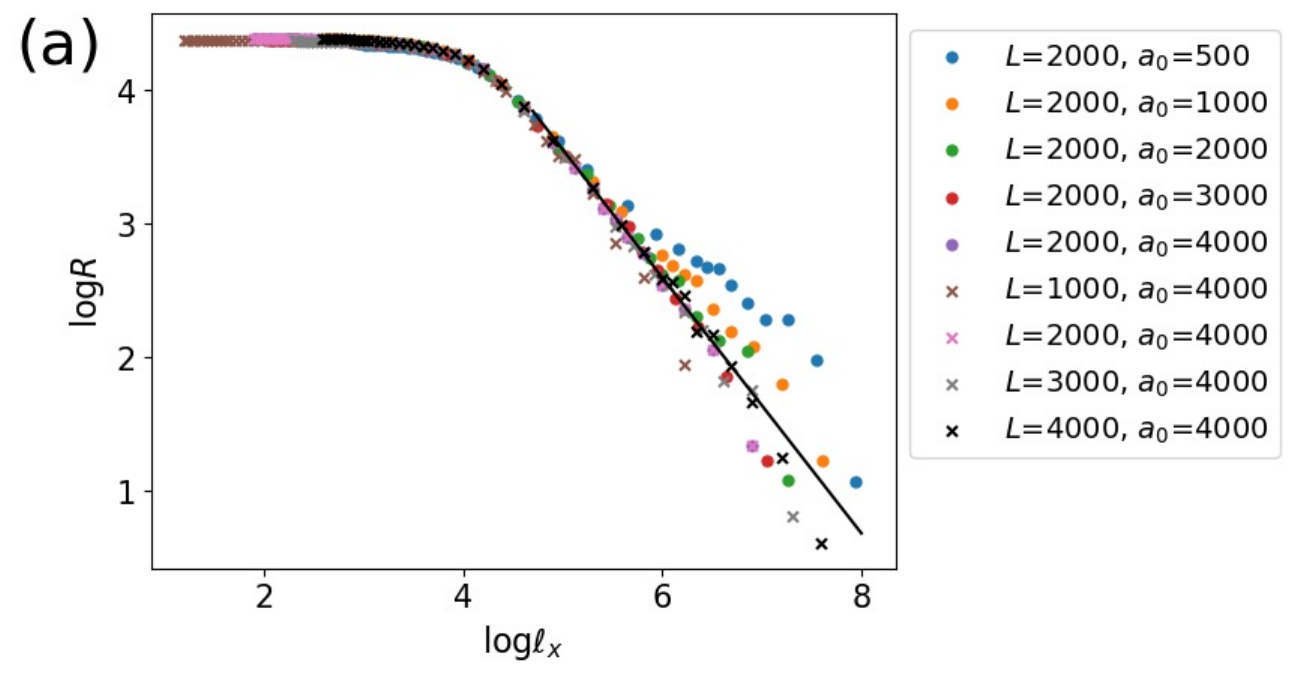}
  \end{minipage}
  \\
\hskip -0.7cm
  \begin{minipage}[b]{0.5\linewidth}
    \centering        
    \includegraphics[keepaspectratio, scale=0.5]    {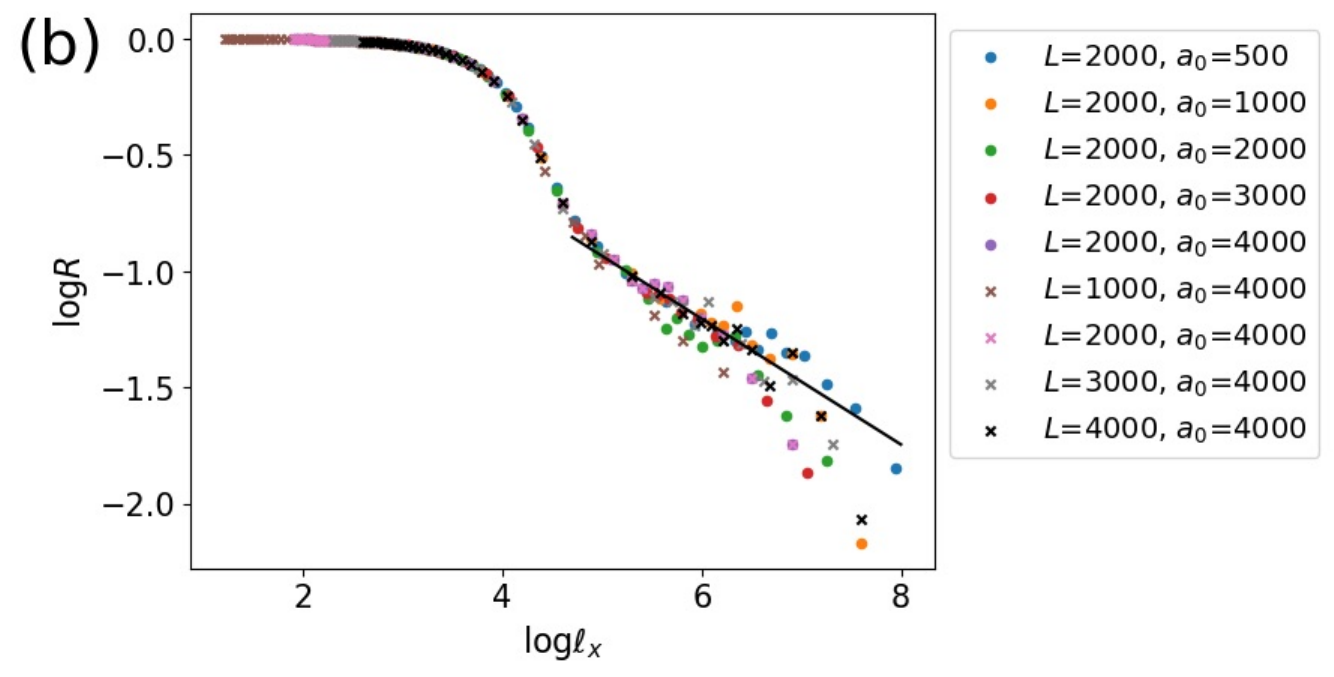}
  \end{minipage}
  \caption{
   $\log R_{\ell_x}$ versus 
    $\log \ell_x$ of the point processes 
    obtained by numerical simulation of the model 
    with variety of system sizes $L=L_x=L_y$ and $a_0$.
(a) For the marked point processes.
    As $L$ and $a_0$ increase, the linear region extends systematically.
    The linear fitting to
    \eqref{eq:fitting} for $L=4000$, $a_0=4000$
    is shown by a black line, whose slope gives
    $-\alpha^{\rm model}_{\rm mass} \fallingdotseq -0.96$.
(b) For the unmarked point processes.
    Even if the values of $L$ and $a_0$ become large, 
    the linear region seems to be restricted
    in a narrow region of scale.
    The linear fitting to \eqref{eq:fitting} 
    in the narrow region for $L=4000$, $a_0=4000$
    is shown by a black line, whose slope gives
    $-\alpha^{\rm model}_{\rm point} \fallingdotseq -0.27$.
}
\label{fig:Landa_0_comp}
\end{center}
\end{figure}

\begin{figure}[ht]
\begin{center}
\includegraphics[width=0.4\textwidth]{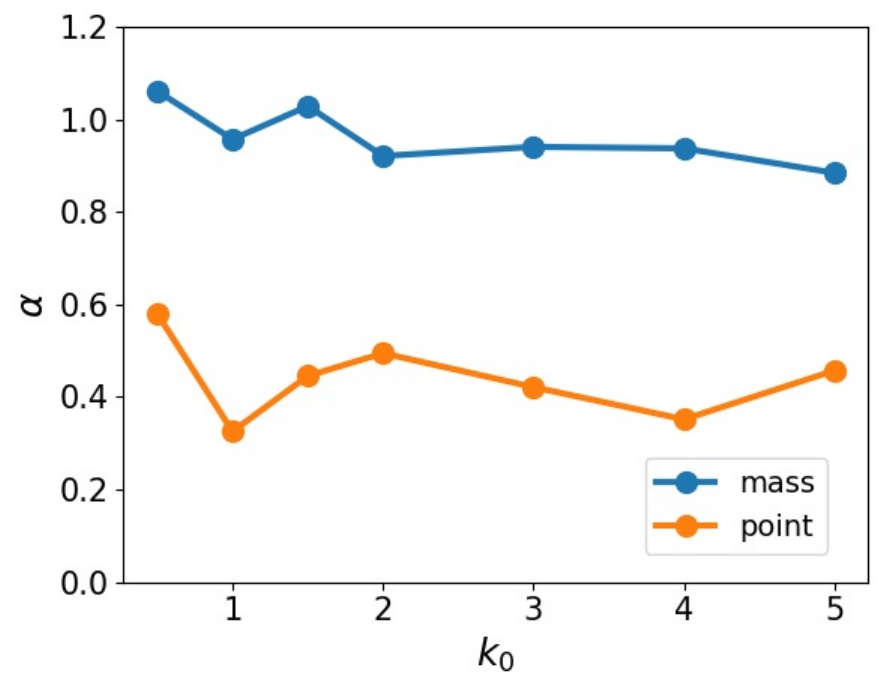}
\end{center}
\caption{
The estimated values of $\alpha$ 
for marked (mass) and unmarked (point) configurations
obtained by the present algorithms 
with different shape parameters
$k_0$ of the Gamma distribution for $A$.
We see that $\alpha^{\rm model}_{\rm mass} \fallingdotseq 1$
and $\alpha^{\rm model}_{\rm point} \fallingdotseq 0.4$. } 
\label{fig:v1.9_k_EvsAlpha}
\end{figure}
Both of the marked point processes
$\Pi$ and the corresponding unmarked point processes
$\Xi_{\Pi}$ obtained by our computer simulations
are systematically studied by changing the system size $L$ and
the mean value $a_0$ of the random area $A$.
Figure \ref{fig:Landa_0_comp} (a) (resp.~(b))
shows 
the $\log R^{\rm mass}_{\ell_x}$ (resp.~$\log R^{\rm point}_{\ell_x}$)
versus $\log \ell_x$ plots
averaged over 20 marked point processes
obtained by our model.
For $500 \leq a_0 \leq 3000$, $L=2000$,
the dots were shifted in the $X \equiv \log \ell_x$ direction
by $\log \sqrt{4000/a_0}$, which 
corresponds to the dilation of the region 
$\Lambda_L$ so that $a_0$ is normalized to be 4000.
In (a) the dots with $500 \leq a_0 \leq 3000$,
$L=2000$ were shifted also in the $Y \equiv \log R_{\ell_x}$ direction
appropriately in order to clarify the convergence of slopes
of lines along which dots line up.

First we explain the results for the marked point processes
shown by Fig.~\ref{fig:Landa_0_comp} (a).
For fixed $L=2000$, as $a_0$ increases from 500 to 4000, 
the linear region extends systematically.
For the largest value $a_0=4000$, 
the linear region also systematically extends
as $L$ increases from 1000 to 4000.
By these results, we think that
if both $L$ and $a_0$ are sufficiently large,
and if the number of iteration-terms $T$ of the transformation $\cT$
is large enough, 
our procedure produced marked point processes
with hyperuniformity.
The linear fitting to \eqref{eq:fitting} of the dots
of $L=4000$ and $a_0=4000$ gives
\begin{equation}
R_{\ell_x}^{\rm mass} \simeq \ell_x^{-\alpha^{\rm model}_{\rm mass}}
\quad \mbox{with} \quad
\alpha^{\rm model}_{\rm mass} \fallingdotseq 0.96.
\label{eq:model_alpha1}
\end{equation}

Next we explain the results for the unmarked point processes.
Figure \ref{fig:Landa_0_comp} (b) shows
the $\log R^{\rm point}_{\ell_x}$ versus $\log \ell_x$ plots
are averaged over 20 unmarked point processes, 
which are obtained from the marked point processes
by deleting the data of mass distributions.
Compared with $\log R^{\rm mass}_{\ell_x}$,
even if the values of $L$ and $a_0$ become large, 
the linear region seems to be restricted
in a narrow region of scale.
The linear fitting \eqref{eq:fitting} to the plots
for $L=4000$ and $a_0=4000$ gives
\begin{equation}
R_{\ell_x}^{\rm point} \simeq \ell_x^{-\alpha^{\rm model}_{\rm point}}
\quad \mbox{with} \quad
\alpha^{\rm model}_{\rm point} \fallingdotseq 0.27.
\label{eq:model_alpha_point}
\end{equation}

We have studied dependence of the results 
on the choice of distribution \eqref{eq:p} of 
random area $A$.
As extensions of the exponential distribution
\eqref{eq:exp_dist}, 
we consider the cases in which the area $A$ follows 
the Gamma distributions with 
the probability density 
\begin{align}
    p(a) 
    = p (a; k_0, \theta_0)
    = \frac{1}{\Gamma(k_0) \theta_0^{k_0}} a^{k_0-1} 
    e^{-a/\theta_0}, 
    \label{eq:Gamma1}
\end{align}
where $\Gamma(k)$ is the Gamma function, 
$\Gamma(k)=\int_{0}^{\infty} t^{k-1} e^{-t} dt$, and
$k_0$ and $\theta_0$ are the shape
and scale parameters, respectively.
It is easy to verify that  
$a_0 \equiv \bra A \ket = k_0 \theta_0$ and
$\Var[A]=k_0 \theta_0^2$.
In particular, when $k_0=1$, \eqref{eq:Gamma1}
is reduced to the probability density of
exponential distribution \eqref{eq:exp_dist}. 
When $k_0 = n \in \N$, the Gamma distribution is called 
the \textit{Erlang distribution} with probability density
$p_0(a; n, \theta_0) =a^{n-1} 
e^{-a/\theta_0}/[(n-1)! \theta_0^n]$, 
and when $k_0=n/2$ with $n \in \N$, 
it is related to the \textit{$\chi^2$ distribution}.
The marked point processes are generated by the 
present algorithms using the Gamma distributions with the shape parameters $k_0=$0.5, 1, 1.5, 2, 3, 4, and 5.
The evaluated values of the exponent $\alpha$ are shown in Fig.$\ref{fig:v1.9_k_EvsAlpha}$. 
The exponent $\alpha$ seems to be
 \begin{align}
 \alpha^{\rm model}_{\rm mass} \fallingdotseq 1, 
 \quad
 \alpha^{\rm model}_{\rm point} \fallingdotseq 0.4.
 \label{eq:alpha_model}
\end{align}

\vskip 0.3cm
\noindent{\bf Remark 2} \,
As a simplified model, we have simulated the algorithm
with a fixed value of area $A$ of disk.
Then we found that the number of iteration-terms $T$
needed for convergence of point processes 
became much larger than 500.
In addition, the convergence of the 
$\log R_{\ell_x}$--$\log \ell_x$ plots
to a line needed larger values of $L$ and $a_0$
compared to the results shown by
Fig.~\ref{fig:Landa_0_comp}.
Hence we have concluded that the randomness of area $A$ of disk
is important to generate marked point processes with
hyperuniformity efficiently.

\subsubsection{Mass distributions}
\label{sec:model_mass}
\begin{figure}[ht]
    \begin{tabular}{cc}
      \begin{minipage}[t]{0.45\hsize}
        \centering
         \includegraphics[keepaspectratio, scale=0.4]
         {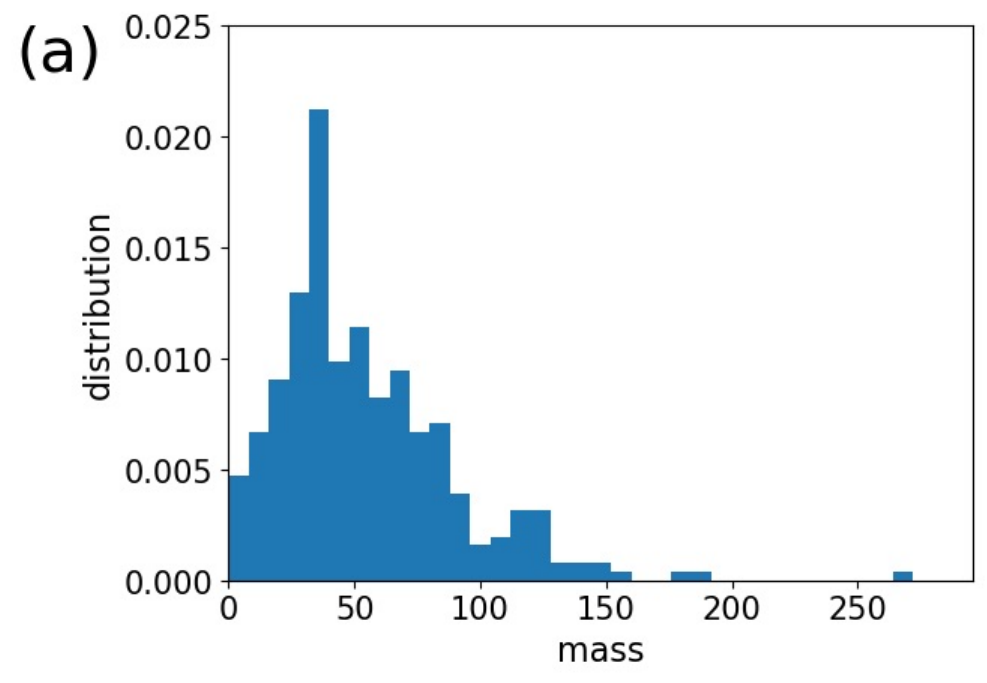}
      \end{minipage} &
      \begin{minipage}[t]{0.45\hsize}
        \centering
         \includegraphics[keepaspectratio, scale=0.4]
         {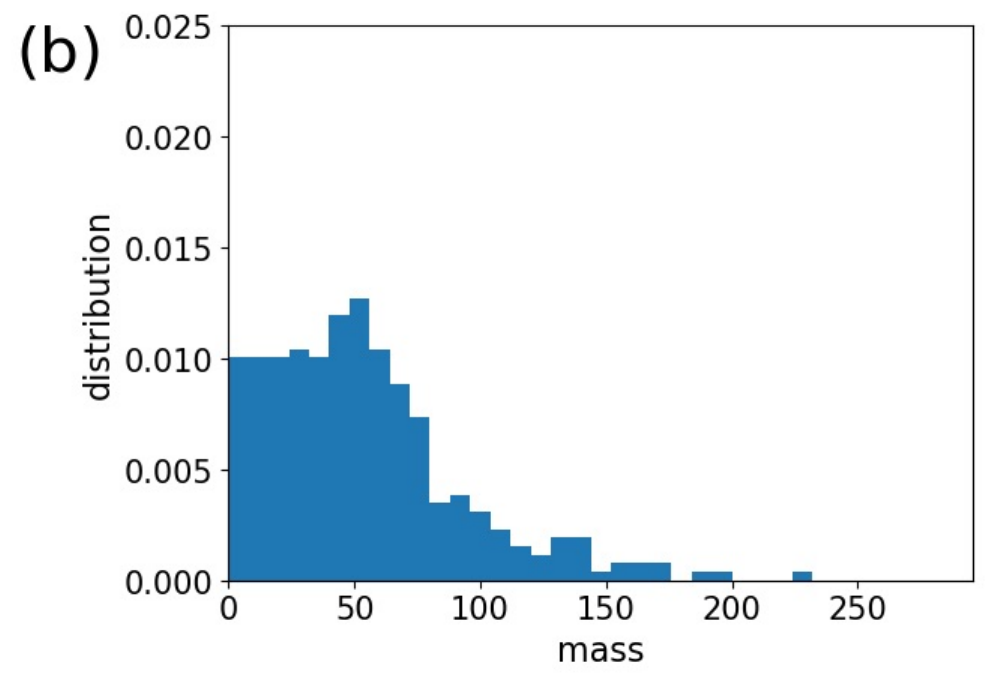}
      \end{minipage}
    \end{tabular}
    \caption{
        Two samples of mass distributions obtained by 
        numerical simulations with $L=2000$ and $a_0=4000$.
        The example (a) shows a peak around 40 in the 
        histogram,
        but the sample (b) shows a plateau under 50. 
        The mass distributions depend on samples.}   
\label{fig:massdist}
\end{figure}

For Figs.~\ref{fig:Landa_0_comp} (a) and (b),
we have used 20 samples 
obtained after $T=500$ iteration-terms of 
transformation $\cT$ starting from independently prepared
20 samples of PPPs.
We found that the mass distributions of 20 samples
seem to be quite different from each other, especially
when $a_0$ becomes large.
Figure~\ref{fig:massdist} shows two samples
of mass distributions obtained by numerical simulations
with $L=2000$ and $a_0=4000$.
The mass distributions depend heavily on samples
and they do not seem to be universal. 

\begin{figure}[ht]
\begin{center}
\includegraphics[width=0.5\textwidth]{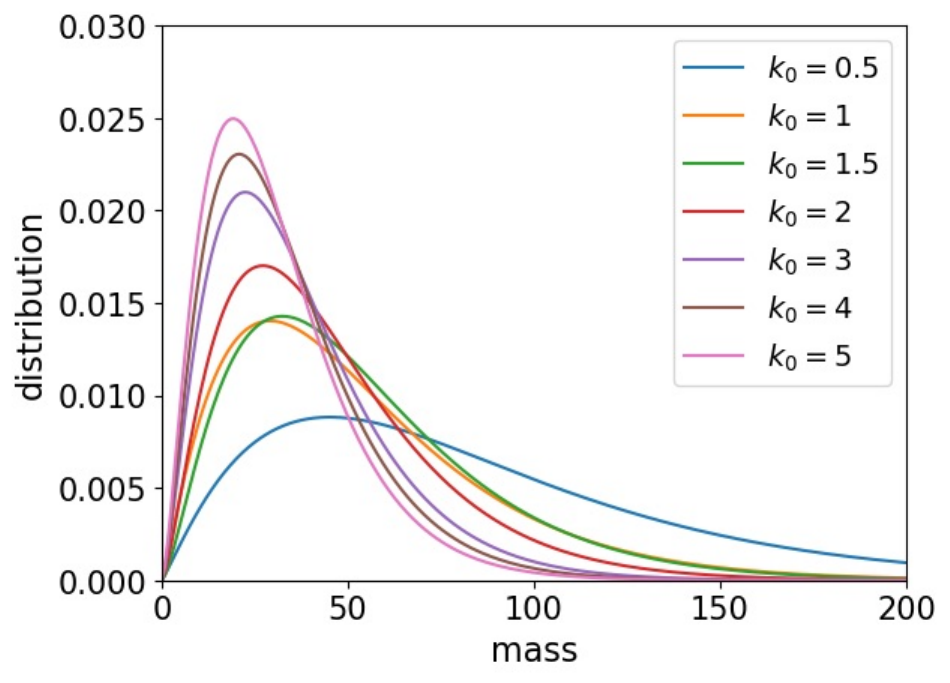}
\end{center}
\caption{
The mass distributions averaged over
the 20 samples are calculated for
several values of $k_0$ of the distribution \eqref{eq:Gamma1}.
Here $L=2000$ and $a_0=k_0 \theta_0=4000$.
They are well-approximated by \eqref{eq:p_G}
with appropriate values of $(k_{\rm G}, \theta_{\rm G})$.}
\label{fig:v1.9_massdist_pdf_All}
\end{figure}

Nevertheless, if we average the mass distributions over 20 samples,
the obtained curves are well approximated by
 the Gamma distributions with appropriate values of parameters
 \begin{align}
    p_{\rm G} (\sm) 
    &= p_{\rm G} (\sm;k_{\rm G}, \theta_{\rm G}) \notag \\
    &= \frac{1}{\Gamma (k_{\rm G}) \theta_{\rm G}^{k_{\rm G}}} 
    \sm^{k_{\rm G}-1} e^{-\sm/\theta_{\rm G}},
    \quad \sm \geq 0.
    \label{eq:p_G}
\end{align}
Here we have fixed the parameters of \eqref{eq:Gamma1} for the 
model as $a_0=k_0 \theta_0=4000$ and $L=2000$.
The $k_0$-dependence of the fitting curves of \eqref{eq:p_G} 
are shown in Fig.\ref{fig:v1.9_massdist_pdf_All}, 
where fitting parameters $(k_{\rm G}, \theta_{\rm G})$ are
given by
$(2.15, 39.3)$, $(2.20, 24.3)$, $(2.50, 21.6)$,
$(2.50, 18.1)$, $(2.55, 14.5)$, $(2.60, 13.0)$, and $(2.60, 12.0)$
for $k_0=0.5, 1, 1.5, 2, 3, 4$, and 5, respectively.

\section{Discussions and Concluding Remarks}
\label{sec:discussion}
In the present paper, we have studied 
hyperuniformity of two groups of marked point processes.
The first group consists of the real samples of bush configurations
weighted by masses observed in deserts.
The second group consists of the numerical samples
of marked point processes 
produced by computer simulations
of our time-evolutionary models, 
in which random thinning of points and 
coalescing of masses are iterated.
For the real data, we have confirmed hyperuniformity
of the marked point processes obtained from
the deserts in Argentina, \textit{bush 1--5}
(see Fig.~\ref{fig:loglog} (b)),
Australia, and Kenya (see Fig.~\ref{fig:other_deserts}).
For the results by computer simulations, hyperuniformity
are evident for the marked point processes
for the models with different choices of
probability distributions of area $A$ used in the
thinning-coalescing processes (see Figs.~\ref{fig:Landa_0_comp} (a)
and \ref{fig:v1.9_k_EvsAlpha}).
In the former, if we do not take into account 
the mass distributions and only regard the real data
as unmarked point processes,
hyperuniformity is not observed (see Figs.~\ref{fig:loglog} (a)).
In the latter, the plots of $\log R_{\ell_x}$ versus $\log \ell_x$
for unmarked point processes $\Xi_{\Pi}$,
which are obtained by deleting the mass information
from marked point processes $\Pi$, do not show
clear power-law decay as shown by
Fig.~\ref{fig:Landa_0_comp} (b).
Moreover, if we look at only mass distributions
of individual points obtained by computer simulations,
they seem to scatter heavily depending on samples
as shown by Fig.~\ref{fig:massdist}. 
These results implies that 
hyperuniformity of marked point processes
can be maintained by 
\textit{strong correlations in probability law}
between spatial configurations of unmarked
point processes and mass distributions of individual points.

As briefly mentioned in Section \ref{sec:three},
hyperuniformity is classified by the values of
the exponent $\alpha$ of power-laws 
(with logarithmic correction in Class II) of 
anomalous suppression of large-scale fluctuations.
The evaluated $\alpha$ for the real data from deserts
are rather scattering from $\alpha^{\rm Kenya} =0.36$
to $\alpha^{\rm Argentina} \fallingdotseq 0.65$, which 
suggests that, if the weighted bush configurations
are in the hyperuniform states,
then they are all in Class III.
On the other hands, the hyperuniform marked
point processes generated by our algorithms
seem to be in Class I with 
$\alpha^{\rm model}_{\rm mass} \fallingdotseq 1$.
The variety of exponent $\alpha$ in the real data
will be due to some geometrical structures.
As a matter of fact, the two samples from 
Algeria are special ones found in a \textit{Wadi}, 
the bed or valley of a stream that is usually dry 
except during the rainy season,
and they do not clearly show hyperuniformity;
$\alpha^{\rm Algeria} \fallingdotseq 0$.

We have succeeded to generate marked point processes
with hyperuniformity from uncorrelated 
PPPs by iterating
random thinning-coalescing processes.
We have no evidence that similar types of thinning and
coalescing processes have been iterated
in continuous survival competitions of bushes
in real deserts. But our numerical simulations
suggest that in order to realize hyperuniformity
in large scale, sufficiently large number of
iterations of local processes should be needed.
As shown by Fig.~\ref{fig:Thinning_comp},
if the number of iteration is not sufficiently large,
evident hyerperuniformity will not be achieved. 
We think that another reason of the
small values of $\alpha^{\rm Kenya}=0.36$ and
$\alpha^{\rm Algeria} \fallingdotseq 0$ will come from
some historical backgrounds of these deserts.

The present work requires the following 
as research subjects in future.
From the view point of ecological study, 
more realistic models and algorithms for bush formations
in deserts shall be considered and 
validity of them should be
tested by real observations. 
We hope fruitful connections of the present study
based on the notion of hyperuniformity of marked point processes
with other important model studies
of vegetation pattern formation,
diversity of ecosystems, and desertification
\cite{Klau99,NT01,vHMSZ01,TY03}.
From the view point of theoretical study
of non-equilibrium statistical physics,
minimal algorithms shall be investigated
to generate marked point processes 
with hyperuniformity
so that the convergence of the algorithms
to highly non-trivial configurations
is able to be proved mathematically.

\vskip 1cm
\noindent{\bf Acknowledgements} \,
AE and MK would like to thank Jun-ichi Wakita
for instructing them the image processing techniques
to produce digital data of bushes in deserts.
AE is grateful to Hiraku Nishimori, 
Mitsugu Matsushita, 
Osamu Moriyama,
Zouha\"ir Mouayn, 
Helmut R. Brand
for useful discussions and encouragement
of her researches. 
MK was supported by JSPS KAKENHI Grant Numbers 
JP19K03674, 
JP21H04432,
JP22H05105,
JP23K25774, 
and
JP24K06888.
TS was supported by JSPS KAKENHI Grant Numbers 
JP20K20884,
JP21H04432, 
JP22H05105,
JP23K25774, 
and 
JP24KK0060.


\end{document}